\PassOptionsToPackage{pdfpagelabels=false}{hyperref}
\documentclass[useAMS,usenatbib]{mnras}
\pdfoutput=1
\usepackage[pdftex]{graphicx}
\usepackage{amsmath}
\usepackage{amssymb}
\usepackage{natbib}
\usepackage{geometry}
\usepackage{rotating,booktabs,multirow}
\usepackage{color}
\usepackage{times}
\usepackage{tabularx}

\newcommand{\bea}{\begin{eqnarray}}
\newcommand{\eea}{\end{eqnarray}}

\volume{{\rm submitted}}

\definecolor{portlandorange}{rgb}{1.0, 0.35, 0.21}
\definecolor{ForestGreen}{RGB}{20, 180, 20}

\newcommand{\Ms}{{\rm M}_\odot}
\newcommand{\MS}{M_{\rm stars}}

\title[Quenched fractions in TNG and quenching pathways]{Quenched fractions in the IllustrisTNG simulations: the roles of AGN feedback, environment, and pre-processing}

\author[Donnari et al.] {Martina Donnari$^{1}$\thanks{E-mail: donnari@mpia-hd.mpg.de}, Annalisa Pillepich$^{1}$, Gandhali D. Joshi$^{1}$, Dylan Nelson$^{2}$, \newauthor 
Shy Genel$^{3}$, Federico Marinacci$^{4}$, Vicente Rodriguez-Gomez$^{5}$, R{\"u}diger Pakmor$^{2}$, \newauthor
Paul Torrey$^{6}$, Mark Vogelsberger$^{7}$, and Lars Hernquist$^{8}$
\\
\\
$^{1}$Max-Planck-Institut f{\"u}r Astronomie, K{\"o}nigstuhl 17, D-69117 Heidelberg, Germany\\
$^{2}$Max-Planck-Institut f{\"u}r Astrophysik, Karl-Schwarzschild-Str. 1, D-85748, Garching, Germany \\
$^{3}$Center for Computational Astrophysics, Flatiron Institute, 162 Fifth Avenue, New York, NY 10010, USA\\
$^{4}$ Department of Physics and Astronomy, University of Bologna, Via Gobetti 93/2, I-40129, Bologna, Italy \\
$^{5}$Instituto de Radioastronom{\'i}a y Astrof{\'i}sica, Universidad Nacional Aut{\'o}noma de M{\'e}xico, A.P. 72-3, 58089 Morelia, M{\'e}xico \\
$^{6}$Department of Astronomy, University of Florida, 211 Bryant Space Science Center, Gainesville, FL 32611, USA\\
$^{7}$ Kavli Institute for Astrophysics and Space Research, Massachusetts Institute of Technology, Cambridge, MA 02139, USA \\
$^{8}$ Harvard-Smithsonian Center for Astrophysics, 60 Garden Street, Cambridge, MA 02138, USA 
}

\begin{document}

\pagerange{\pageref{firstpage}--\pageref{lastpage}} \pubyear{2020}

\maketitle

\label{firstpage}

\maketitle

\begin{abstract}
We use the IllustrisTNG hydrodynamical simulations to show how the fractions of quenched galaxies vary across different environments and cosmic time, and to quantify the role AGN feedback and pre-processing play in quenching group and cluster satellites. At $z=0$, we select galaxies with $\MS = 10^{9-12} ~ \Ms$ residing within ($\leq R_{200c}$) massive groups and clusters of total host mass $M_{200c} = 10^{13-15.2} ~ \Ms$ in TNG100 and TNG300.
The model predicts a quenched fraction of $\sim$70-90 per cent (on average) for centrals and satellites of mass $\gtrsim 10^{10.5} ~ \Ms$, regardless of host mass, cosmic time ($0\leq z\leq0.5$), cluster-centric distance and time since infall in the $z=0$ host. 
Low-mass central galaxies ($\lesssim 10^{10} ~ \Ms$), on the other hand, are rarely quenched unless they become members of groups ($10^{13-14} ~ \Ms$) or clusters ($\geq10^{14} ~ \Ms$), where the quenched fraction rises to $\sim$ 80 per cent.
Typically, the fraction of low-mass passive galaxies is higher closer to the host center and for progressively more massive hosts. The population of low-mass satellites accreted more than $\sim$ 4-6 Gyr ago in massive hosts is almost entirely passive, thus suggesting an upper limit for the time needed for environmental quenching to occur.
In fact, about 30 per cent of group and cluster satellites that are quenched at $z=0$ were already quenched before falling into their current host, and the bulk of them quenched as early as 4 to 10 billion years ago. 
For low-mass galaxies ($\MS \lesssim10^{10-10.5}~ \Ms$), this is due to \textit {pre-processing}, whereby current satellites may have been members of other hosts, and hence have undergone environmental processes, before falling into their final host, this mechanism being more common and more effective for the purposes of quenching for satellites found today in more massive hosts. On the other hand, massive galaxies quench on their own and because of AGN feedback, regardless of whether they are centrals or satellites.

\end{abstract}


\section{Introduction}
\label{intro}

Galaxies in the Universe can be broadly classified into two main categories: ``passive'' or ``quenched'' galaxies, in which the star formation has ceased, and ``star-forming'' galaxies characterized by ongoing star formation. Typically, the first population exhibits spheroidal or elliptical morphologies and is mostly observed at high stellar masses or populating the high-density environments, while the latter preferentially shows disk-like morphologies and resides in the low-density regions of the Universe \citep{2001Strateva,2003Kauffmann,2004Baldry}.

Observations and theoretical models agree on the bimodal distribution in the nearby Universe of many galaxy properties such as color, morphology, and star formation rate (SFR), although the bimodal nature of the latter is still a matter of debate \citep[see e.g.][]{2017Feldmann,2018Eales}. Furthermore, it is widely accepted that, given the hierarchical nature of the growth of structures, galaxy properties are tightly linked to whether a galaxy is a central or a satellite of its underlying dark matter host halo.

In the last few decades, several authors have proposed a wide variety of quenching mechanisms capable of describing the observed properties of galaxies. Although these mechanisms are still not completely understood, the cessation of star production in galaxies can be divided into two main scenarios, widely known as ``mass quenching'' and ``environmental quenching'', since both the galaxy stellar mass and the environment in which galaxies reside and evolve have been shown to play a crucial role in the star formation activity of galaxies \citep[e.g.][]{2010Peng,2012Vulcani,2017Darvish,2016paccagnella}.

From observations, galaxies more massive than a few $10^{10} \, \Ms$ are likely to be quenched even if they reside in low-density regions. In this case, the quenching mechanisms are generally associated to the internal evolution of these galaxies, although galaxy mergers are also invoked as indirect quenching mechanisms \citep{2009VanDerWel}, leading, for example, to gas consumption through starbursts. Among the ``secular'' processes, from a theoretical perspective, it is widely accepted that in massive galaxies, feedback from Active Galactic Nuclei (AGN) plays an important role in regulating star formation and in quenching galaxies
\citep[e.g.][]{2005DiMatteo,2005Hopkins,2013Silk,2014Hopkins,2015Somerville,2017Beckmann,2018Penny}. In fact, numerical cosmological simulations and semi-analytic models that follow galaxy evolution with and without AGN feedback \citep[e.g.,][]{2011McCarthy,2016Dubois,2017Kaviraj,2018Weinberger} have demonstrated that broad consistency with observations with respect to the fractions of quenched galaxies and galaxy color distributions is only possible when some form of AGN feedback is included.

Conversely, a trend of lower SFRs in higher density environments, such as groups and clusters of galaxies, has emerged from observations of low-mass galaxies ($\lesssim 10^{10} \, \Ms$), while the latter are rarely quenched if isolated \citep[see e.g.][]{2012Geha,2018SAMI}. For example, recently, \cite{2019Schaefer} have shown that the specific Star Formation Rate (sSFR = SFR/$\MS$) of SAMI (Sydney AAO Multi-object Integral Field Galaxy Survey) galaxies drops as the local environment density increases around them.
The effect of group environment has also been extensively investigated in SDSS data; for example, more recently in \citealt{2017Crossett}, by comparing (NUV-r) colors of both satellites and centrals: the authors find that a residual SFR is preferentially seen in field galaxies compared to satellites in the same stellar mass range. 
Various physical processes are commonly invoked to explain quenching driven by the local environment, where galaxies might be subjected to many mechanisms, some of them capable of reducing their gas reservoirs. Among others, these include strangulation or starvation \citep{1980Larson,1999moore,2011Nichols,2015Peng}, tidal stripping \citep{1983Merritt}, galaxy harassment \citep{1996Moore}.  Furthermore, one of the processes often proposed for satellite galaxies is ram pressure stripping \citep{1972Gunn,1999moore,2017Poggianti,2018Barsanti}: as the gas is lost into the intra cluster-medium (ICM), the amount available to produce stars inside the galaxy gradually decreases, eventually leading the star formation to cease. This mechanism has been declared as the main contributor to quench observed galaxies in massive clusters, such as in Virgo \citep[see e.g.][and references therein]{2014Boselli,2016Boselli}, and can now also be investigated in great detail in modern cosmological hydrodynamical galaxy simulations \citep[see e.g.][for a census of jellyfish galaxies in the IllustrisTNG simulations]{2019Yun}.

The emergence of two distinct quenching scenarios is expected to produce different fractions of quenched galaxies when taking into account galaxies in isolation rather than satellites in groups and clusters. 
The fraction of quenched galaxies (all together, i.e. without distinguishing between centrals and satellites) has been shown to increase with increasing galaxy stellar mass at any redshift \citep[$z\lesssim3$, see e.g.][]{2013Muzzin,2016darvish,2018Jian,2017Fossati,2018Fang}. When only satellite galaxies are considered, the quenched fraction is found to be related to the host mass in which satellites reside, increasing with increasing host-to-satellite mass ratio, in observations \citep{2012Wetzel,2013Wetzel,2019Davies} as well as in simulations \citep{2015Furlong,2017Bahe,2019Tremmel} and in semi-analytic models \citep[SAMs;][]{2012Delucia,2019Delucia,2017Henriques}. Furthermore, satellites are more likely to be quenched if they were accreted early, reinforcing the idea of an environmentally-driven quenching scenario \cite[see e.g.][]{2019Pasquali,2017Rhee}. 
Finally, the role of the environment is also found to be strongly correlated with the position of galaxies within their hosts: satellites are more likely to be quenched if they reside towards the cluster center rather then in the outskirts, this trend being stronger for lower-mass satellites  \citep{2012presotto,2013Wetzel,2018Barsanti}.

Besides the most massive clusters (total mass $\gtrsim10^{14}\Ms$), lower-mass groups are widely recognized to play a fundamental role in shaping the evolution of their satellite galaxies, as they are the building blocks of hierarchical cosmic structure formation in the framework of the $\Lambda$CDM model.
In fact, it is widely accepted that a fraction of galaxies that today reside in massive groups and clusters have been orbiting in smaller subgroups prior to their accretion onto their current host, experiencing what is generally known as ``pre-processing'' \citep{1996Zabludoff,2004Fujita,2009Mcgee,2014Hou,2019Joshi,2019Bahe}. However, albeit this mechanism has been widely investigated in observations \cite[e.g. in the LoCuSS survey and SDSS:][and references therein]{2015Haines,2018Bianconi,2013Wetzel} as well as in theoretical models \citep[analytical models, semi-analytical models and cosmological hydrodynamical zoom-in simulations of galaxy clusters:][]{2009Mcgee,2013Bahe,2019Bahe,2017Rhee,2017Joshi,2018Han}, its importance in determining galaxy quenched fractions is still unclear. 

For example, observationally, \cite{2014Hou} have investigated the quenched fractions of $0.01<z<0.045$ satellites in SDSS groups and clusters. By means of a combination of simulations and mock catalogues, they inferred that pre-processing plays a significant role in establishing the observed quiescent fractions, but only when considering the most massive clusters in their sample ($> 10^{14.5} \, \Ms$). These results are somewhat consistent with results from SAMs described in \cite{2012Delucia} and \cite{2009Mcgee}, in which $\sim$ 25-45 per cent of their simulated cluster galaxies were first accreted by subgroups and then fell into the current host. Similar conclusions have been outlined in \cite{2013Bahe,2019Bahe} by using \texttt{GIMIC} (a set of re-simulations of approximately spherical regions, extracted from the Millennium Simulation, which include $\sim$100 hosts with masses in the range $M_{200c}=10^{13-15.2}\,\Ms$) and Hydrangea (a suite of  cosmological hydrodynamical zoom-in simulations of massive galaxy clusters, $M_{200c}=10^{14-15}\,\Ms$), respectively, arguing that the frequency and impact of pre-processing strongly depends on the host mass at the time of observation: more massive clusters have higher fractions of pre-processed galaxies compared to lower-mass groups. Similar conclusions on the dependence on host final mass of the fraction of pre-processed galaxies have been obtained via an independent analysis of the \texttt{EAGLE} simulation by \cite{2019Pallero}. On the other hand, \cite{2009Berrier}, using cosmological $\Lambda$CDM N-body simulations, have previously argued that pre-processing cannot significantly contribute to the present-day quenched fractions in clusters, since the majority of their satellites fell directly into their current host.

Differently from observations, the possibility of actually following the evolution of galaxies across cosmic time back to their infall into clusters represents the power of theoretical models. Yet, the fact that cluster galaxies might have been orbiting in smaller groups prior to being accreted in their final host does not automatically imply that their star formation activity has been impacted in such earlier stages to the point of quenching: this connection can be probed only with numerical models that self-consistently follow all the mechanisms that are thought to be relevant for galaxy evolution within the full cosmological context.

One such example is represented by the state-of-the art cosmological magneto-hydrodynamical simulations of \textit{The Next Generation} Illustris project \citep[hereafter IllustrisTNG or TNG:][]{2018Naiman,2018Marinacci,2018Springel,2018Pillepich,2018Nelson,2019Pillepich_50,2019Nelson_50}. These sample tens of thousands of galaxies orbiting in hundreds of groups and clusters, thus providing one of the most reliable laboratories to study the evolution of galaxies across different environments and cosmic time. Importantly, unlike in most zoom-in suites, galaxies and halo hosts in the TNG volumes represent unbiased samples of the average galaxy and halo mass distributions in the Universe, thus allowing us to derive statistically-robust and unbiased relationships across the mass spectrum. 

In this paper we aim to leverage the outcome of the IllustrisTNG simulations to pinpoint all the various pathways towards galaxy quenching by accounting for the effects of the hierarchical growth of structure, and hence by establishing the role of environment and pre-processing in determining the star-formation activity of galaxies as a function of galaxy and host mass. In practice, we will address the following questions: What is the role of pre-processing in quenching galaxies prior to their accretion onto their $z=0$ host?
What is the distribution of the host masses where they actually quench? And since when are $z=0$ satellites quenched? Do high-mass groups and clusters quench their satellites more efficiently with respect to lower-mass groups even when pre-processing is taken into account? And do massive satellites quench because of environmental or internal processes?

Several outcomes from the IllustrisTNG simulations have validated the model against observational constraints, making them suitable for the tasks at hand. Among them we highlight: the shape and width of the red sequence and the blue cloud of $z=0$ galaxies \citep{2018Nelson}, the existence and locus of the star formation main sequence at low redshifts \citep{2019Donnari}, the distribution of stellar mass across galaxy populations at $z\lesssim4$ \citep{2018Pillepich}, the galaxy size-mass relation for star-forming and quiescent galaxies at $0\le z \le 2$ \citep{2018Genel}, the evolution of the galaxy mass-metallicity relation \citep{2018Torrey}, and quantitatively consistent optical morphologies in comparison to Pan-STARRS data \citep{2019Rodriguez-Gomez} as far as galaxy properties are concerned. These come in addition to observationally-consistent results in relation to the properties of massive hosts and their intra-halo gas: e.g. the X-ray signals of the hot gaseous atmospheres \citep{2020Truong, 2020Davies}, the amount and distribution of highly-ionized Oxygen around galaxies \citep{2018NelsonB}, and the distributions of metals in the intra-cluster medium at low redshifts \citep{2018Vogelsberger}.
Importantly, the new physical mechanisms included in the TNG model have been shown to return galaxy populations whose star formation activity is in better agreement with observations than previous calculations like Illustris \citep{2014vogel,2014Genel,2015Sijacki}, in galaxy colors \citep{2018Weinberger,2018Nelson}, atomic and molecular gas content of satellite galaxies \citep{2019Stevens}, and quenched fractions of central and satellite galaxies taken together with no distinction \citep[$z\lesssim2$,][]{2019Donnari}. In a companion paper (\cite{Donnari2020b}, we discuss in detail the level of agreement between TNG results and observations when centrals and group and cluster satellites are considered separately and show that the quenched fractions of TNG galaxies are overall consistent with observational constraints and hence the results of this paper trustworthy.

This paper is organized as follows. 
In Section \ref{method} we introduce the IllustrisTNG simulations and describe the criteria and definitions used throughout.
In Section \ref{results} we demonstrate the main results from the TNG model: we investigate the fraction of quenched galaxies versus galaxy stellar mass of both centrals and satellites, across cosmic time. We also present our findings across environments, i.e. host masses for satellite galaxies, and explore how the quenched fractions are correlated with the $z=0$ cluster-centric distance, in concert with its distribution in phase space at $z=0$ and its dependence on the amount of time spent in group and cluster environments.
In Section \ref{discussion} we introduce the concept of pre-processing and we discuss its role in shaping the $z=0$ quenched fraction of satellites, show the distributions of the mass of their quenching host and the time since quenching, and draw a picture of all the different pathways galaxies quench according to the IllustrisTNG simulations.
Our summary is given in Section \ref{summary}.

\section{Methods and definitions}
\label{method}

\subsection{The IllustrisTNG simulations}

In this work, we use two simulations modelling two cosmological volumes with, respectively, a side length of 75 $h^{-1}\approx$ 100 Mpc and a baryonic target mass of $m_{\rm baryon}\sim 1.4\times 10^6 \Ms$ (TNG100) and 205 $h^{-1}\approx$ 300 Mpc a side and a baryonic target mass of $m_{\rm baryon}\sim 1.1\times 10^7 \Ms$ (TNG300).
These are part of the IllustrisTNG project \footnote{\url{http://www.tng-project.org}} \citep{2018Naiman,2018Marinacci,2018Springel,2018Pillepich,2018Nelson,2019Pillepich_50,2019Nelson_50}.
The highest resolution simulation of the project, with a side length box of 35 $h^{-1}\approx$ 50 Mpc and a baryonic target mass of $m_{\rm baryon}\sim 8.5\times 10^4 \Ms$ \citep[TNG50: ][]{2019Pillepich_50,2019Nelson_50}, will not be discussed in this paper.

The TNG100 and TNG300 runs are cosmological magneto-hydrodynamical simulations performed with the \texttt{AREPO} code \citep{2010springel} and including various physical ingredients, such as a novel black hole feedback scheme. All details of the underlying galaxy formation model can be found in the two methods papers by \cite{2017Weinberger} and \cite{2018Pillepich}; further details about the numerical resolution achieved in TNG100 and TNG300 can be found for example in \cite{2018Pillepich,2018Nelson}.
All the results presented in this paper are the direct outcomes of TNG100 and TNG300: we do not rescale any property of the simulated galaxies to account for resolution effects, as proposed in previous works \citep{2018Pillepich, 2020Engler}. Instead, we will comment throughout and in Appendix~\ref{appendix} on how numerical resolution affects the presented results.
\subsection{Galaxy and cluster samples: centrals versus satellites}
\label{sec:sample}

\begin{table*}
\centering
\begin{tabular}{l|c|c|c|c}
\hline
\multirow{3}{*}{} & \multicolumn{2}{|c}{TNG100} & \multicolumn{2}{|c}{TNG300}  \\    
\cline{2-5}
& All galaxies & Cluster galaxies & All galaxies & Cluster galaxies \\
\cline{2-5}
& $\MS>10^9 ~\Ms$ & $\MS >10^9~\Ms$ & $\MS>10^9 \,\Ms$ & $\MS>10^9 ~\Ms$ \\
&                  & $M_{200c} = 10^{13-14.6} ~\Ms$ &  & $M_{200c} = 10^{13-15.2} ~\Ms$  \\

\hline
\hline
$z=0$ Centrals   & 10942 & 182  & 149006 & 3733  \\
$z=0$ Satellites ($<R_{200c}$) & 4595  & 2857 & 60184  & 39540  \\
\hline         
\end{tabular}
\caption{\label{tab:sample} Number of satellites and centrals in TNG100 and TNG300 at $z=0$. The label ``all galaxies'' denotes the whole population while ``cluster galaxies'' indicates only satellites residing in groups and clusters, as explained in the text. In both TNG100 and TNG300, the number of central galaxies in groups and clusters is equal to the number of hosts contained in each simulation. The indicated stellar mass $\MS$ and the virial host mass $M_{200c}$ are the minimum masses above which we gather our samples.}
\end{table*}

All the properties of haloes and subhaloes used in this work are obtained via the Friends-of-Friends (FoF) \citep{1985Davis} and \texttt{SUBFIND} algorithms \citep{2001springel,2009Dolag}, which are used to identify substructures across the simulations. 

In order to investigate the effect of environment on galaxy evolution, we differentiate galaxies residing in groups and clusters from the whole population. 
Unless otherwise stated, we define a ``virial'' radius $R_{200c}$ as the radius within which the mean enclosed mass density is 200 times the critical density of the Universe. We will refer to the total mass enclosed within this radius as $M_{200c}$. 
In general, all haloes with $M_{200c} \geq 10^{13} \, \Ms$ will be named ``hosts''. 
More specifically, throughout the text we will refer to hosts less massive than $10^{14} ~\Ms$ as ``groups'', and more massive ones as ``clusters''.

Furthermore, at $z=0$, all galaxies residing within one virial radius of the host center are dubbed as ``satellites'' but for the galaxy settled at the  minimum of the potential well of its FoF halo, which is typically the most massive and will be named the ``central''.
Galaxies in hosts with $M_{200c} \geq 10^{13} \, \Ms$ are generally called group or cluster galaxies.
Moreover, we neglect all those subhalos that are not thought to be of cosmological origin, i.e. they are possibly baryonic fragments of other galactic structures identified by the Subfind algorithms \citep[see Section 5.2 in][]{2019Nelson_release}.
We note here that, for the purposes of the paper, the adjective ``satellite'' assumes a more general meaning at $z>0$: namely, when we track cluster galaxies back in time, we also define as satellites those galaxies orbiting outside the virial radius $R_{200c}$ of their $z=0$ host.

At $z=0$, TNG300 (TNG100) samples 3733 (182) among groups and clusters of mass $M_{200c} \geq 10^{13} \, \Ms$ and a total of more than 39,000 (2,800) satellites with stellar mass $\MS\geq 10^9\Ms$. The most massive cluster in TNG300 (TNG100) reaches a total host mass of $10^{15.2} ~ \Ms$ ($10^{14.6} ~ \Ms$) -- see Tab. \ref{tab:sample} for more details.

The galaxy stellar mass $\MS$ adopted in this work is the sum of all gravitationally bound stellar particles within twice the stellar half-mass radius, $R_{\rm star,h}$.
All galaxies in our samples have a stellar mass larger than $\MS \geq 10^9 ~ \Ms$, that corresponds to only having galaxies in our sample with at least one thousand star particles in total in TNG100 and more than one hundred in TNG300.
Throughout the text and in all the figures presented in this work, we will clarify whether we take into account satellites, centrals or both.

\subsection{Star formation rates in IllustrisTNG galaxies}
For the purposes of this work we characterize galaxies by their `instantaneous' SFR, i.e. by considering the sum of the individual SFR of all gas cells within $2\times R_{\rm star,h}$.
The effects of different averaging timescales and of the aperture choice for the SFR of TNG galaxies has been extensively investigated in \cite{2019Donnari},  where we have shown that the former (latter) are (not) negligible for the identification of the locus of the star-forming main sequence (MS). In the companion paper, \cite{Donnari2020b}, we emphasize that the aperture choice has non-negligible effects on the values of galaxies' SFR, particularly when galaxies are being quenched, and hence any quantitative comparison to the results of this paper must take this notion into account. 
We note that even if all galaxies in our sample are well-resolved in terms of stellar and gas resolution elements, this is not the case for their SFR values. We assign an SFR value extracted randomly in the range $(10^{-5} - 10^{-4}) \, \Ms \rm yr^{-1}$ to those galaxies with Log SFR below the resolution limit of the simulation, which otherwise would have SFR$\equiv$0.


\subsection{Quenched versus star-forming galaxies}
\label{QvsSF}

Throughout the text, we consider the adjectives ``quenched'', ``quiescent'', and ``passive'' as having interchangeable meanings and we do not distinguish between star-forming and star-burst galaxies, nor between quiescent and green-valley galaxies. In \cite{2019Donnari} and \cite{Donnari2020b}, we explore and compare a number of different criteria to separate quenched versus star-forming galaxies; here we use the following definition throughout the study.

{\it Quenched galaxies are those whose SFR is one dex below the star-forming main sequence.} 
Following the same procedure introduced in \cite{2019Donnari} and \cite{2019Pillepich_50}, we stack all galaxies in 0.2 dex bins of stellar mass, measure the median SFR in the bin, and name as ``quenched'' those galaxies whose SFR falls below 1 dex from the median SFR at the corresponding mass. 
We then remove the quenched galaxies, calculate a new median, and repeat these steps iteratively until the median SFR in the mass bin is converged to a given accuracy.
For this paper we have chosen a convergence criterion where the median SFR in successive iteration steps differ by less than 1 per cent.
This allows us to find the locus of the star-forming Main Sequence (MS), that in TNG is linear in the $\MS=10^{9-10.2} ~ \Ms$ range and can be fitted using Eq. (1) in \cite{2019Donnari}. Galaxies whose SFR is 1 dex lower than the fitted MS at the corresponding stellar mass are flagged as quenched; for galaxies more massive than $\MS=10^{10.2} ~ \Ms$ the linear extrapolation of the MS is used as reference. 

We emphasize that the choice of extrapolating the locus of the MS beyond a certain mass and the precise stellar-mass range considered to fit the MS may have a non-negligible impact in the quantitative assessments of the quenched fractions, depending on redshift and particularly in regimes with low number statistics: we estimate the systematic errors associated with the latter choice to be as large as 10-20 percentage points (see Fig. 3 in \citealt{2019Donnari} and \citealt{Donnari2020b}).
On the other hand, we note that, at low redshift, the separating criterion based on the MS used in this work is equivalent (to better than a few percentage points in quenched fraction) to a fixed sSFR $\leq 10^{-11} \, \rm yr^{-1}$ cut which is often used in low-$z$ environmental observational studies \cite[Fig. 1 of][]{Donnari2020b} and at $z\le2$ to UVJ cuts \cite[Fig. 3 bottom panels of][]{2019Donnari}. Therefore, we estimate that the majority of the results in the following (those at $z\lesssim1$) are quantitatively robust to better than 10 per cent against quenched definition and we comment accordingly otherwise.

\section{Results from the TNG simulations}
\label{results}

\subsection{Fractions of quenched galaxies: centrals versus satellites at $z=0$}
\label{sec:qfrac}

\begin{figure*}
\centering
\includegraphics[width=0.48\textwidth]{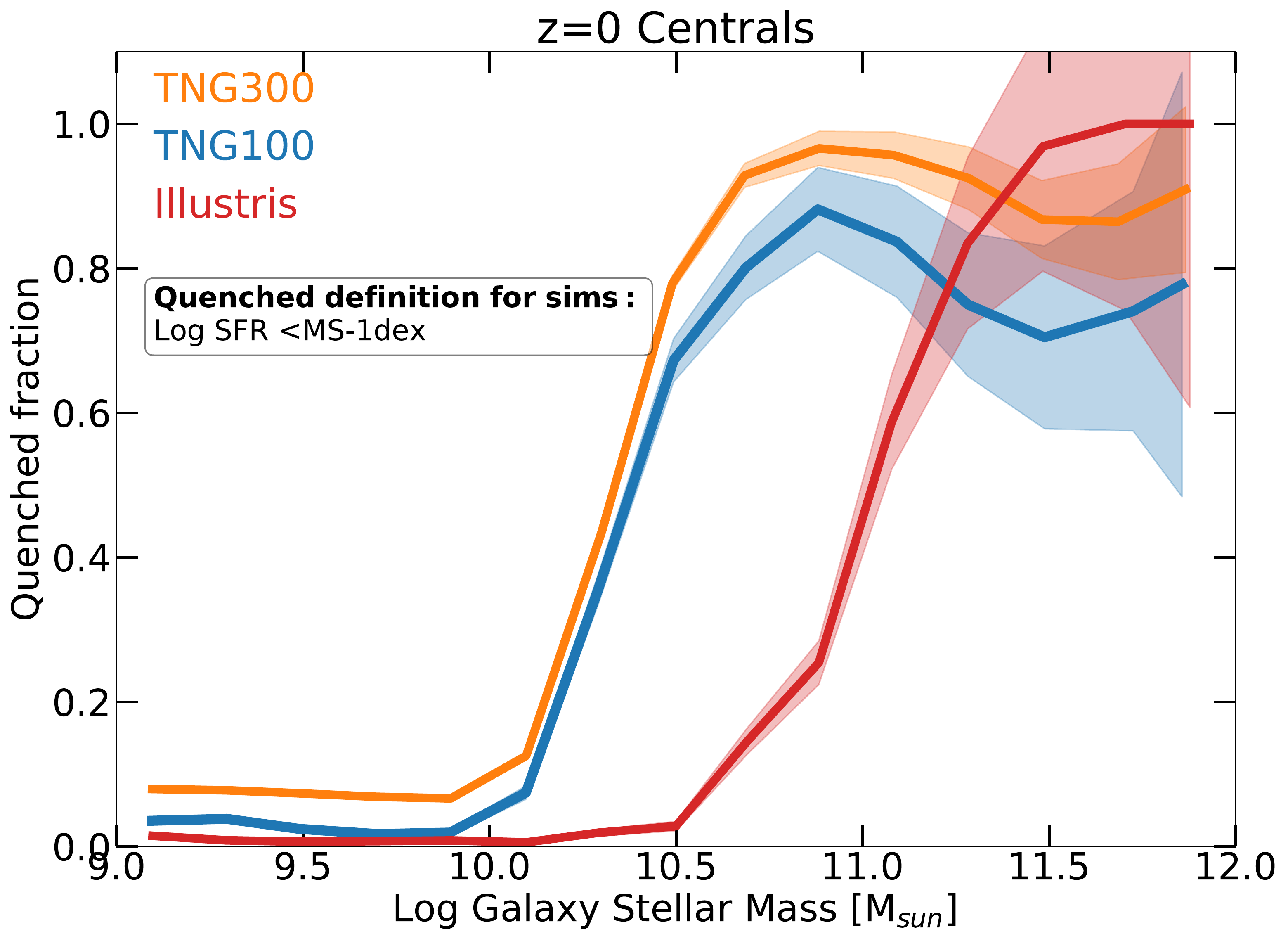}
\includegraphics[width=0.48\textwidth]{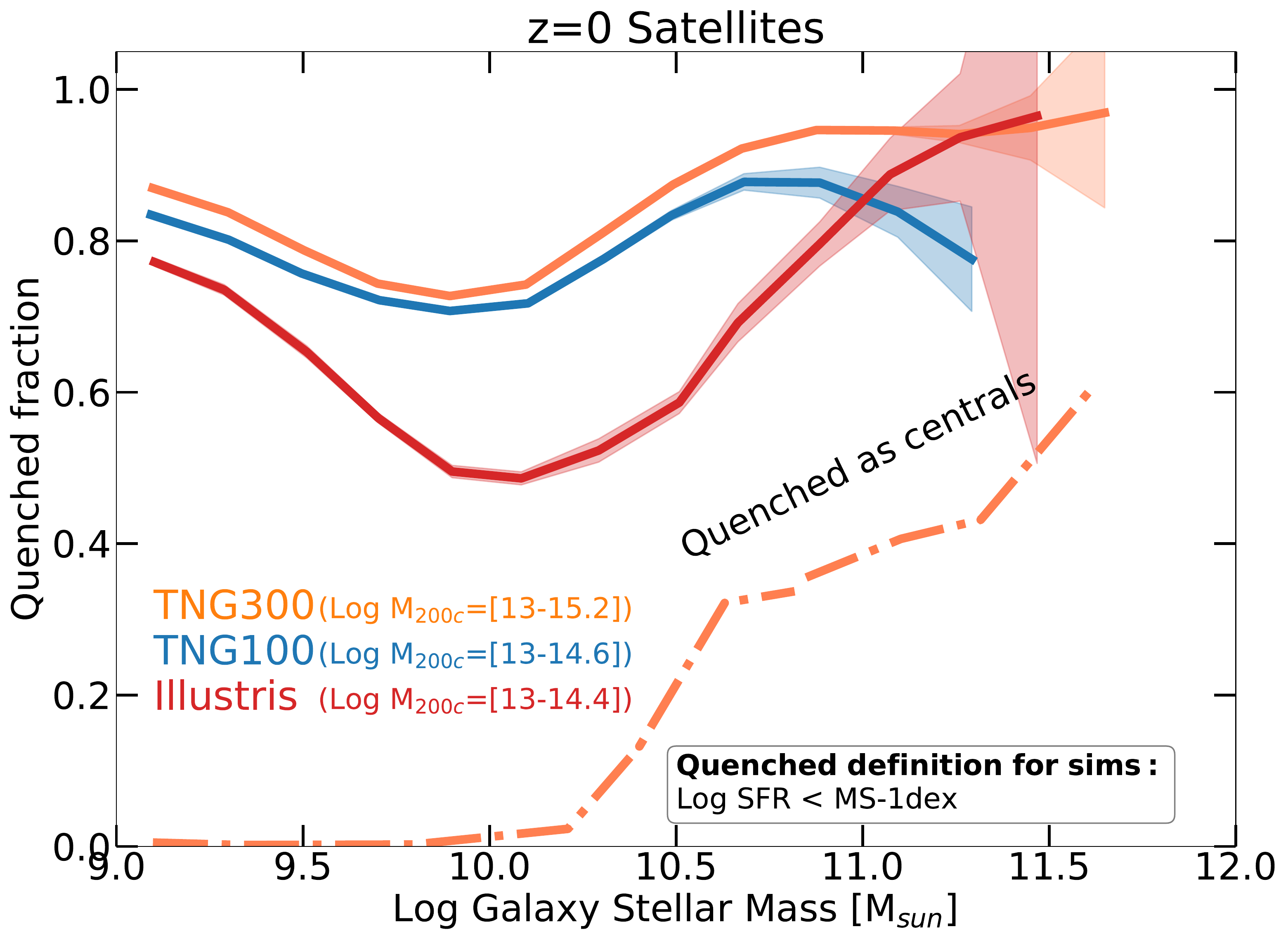}
\caption{\label{fig:Q_frac} {\bf IllustrisTNG quenched fractions at $z=0$}. Fraction of quenched central (left panel) and satellite (right panel) galaxies at $z=0$ for TNG300 (orange), TNG100 (blue) and Illustris (red), in bins of 0.2 dex in stellar mass. Satellites are galaxies within the 3D $R_{200c}$ of their host, excluding the central.
We use the quenched definition of $\rm Log \, SFR < MS$ -1dex here. The dash-dotted curve in the right panel represents the quenched fraction for satellites prior to any infall. At the low-mass end, the quenched fractions are significantly higher for satellites in groups and clusters compared to centrals of similar stellar mass.}
\end{figure*}

In this section, we investigate the fractions of quenched galaxies as a function of galaxy stellar mass at $z=0$, according to the TNG model. Our findings are presented in Fig. \ref{fig:Q_frac}.

We evaluate the $z=0$ quenched fraction separately for centrals (left panel) and satellites (right panel) in TNG300 (orange curve) and TNG100 (blue curve), and we compare the results with the original Illustris simulation (red curve), to better highlight the different efficiencies of the quenching mechanisms among these two numerical models. We consider all satellites in the simulated volumes that reside in hosts at least as massive as $10^{13}~\Ms$: hence the TNG100 and TNG300 curves represent somewhat different satellite populations as in the latter, satellites can be hosted by more massive haloes, up to $\sim 10^{15.2}~\Ms$. We plot the raw data, i.e. galaxy stellar masses and SFR values taken directly from the simulation output without any correction. We define quenched galaxies as those having SFR 1 dex below the MS.
From now on, we only present results if there are at least 10 galaxies per bin. Shaded areas denote the Poissonian error per stellar mass bin.

\begin{figure*}
\centering
\includegraphics[width=0.48\textwidth]{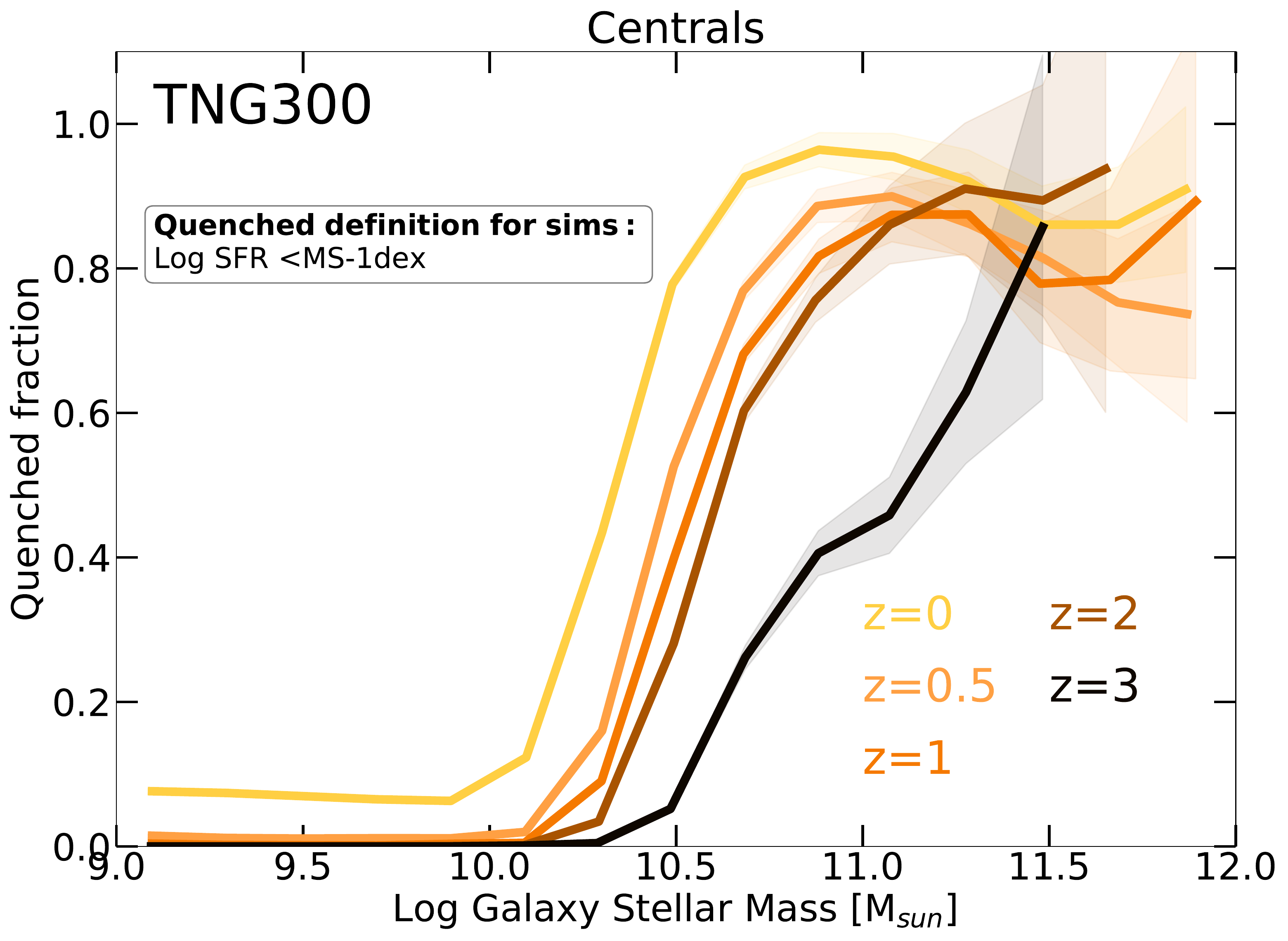}
\includegraphics[width=0.48\textwidth]{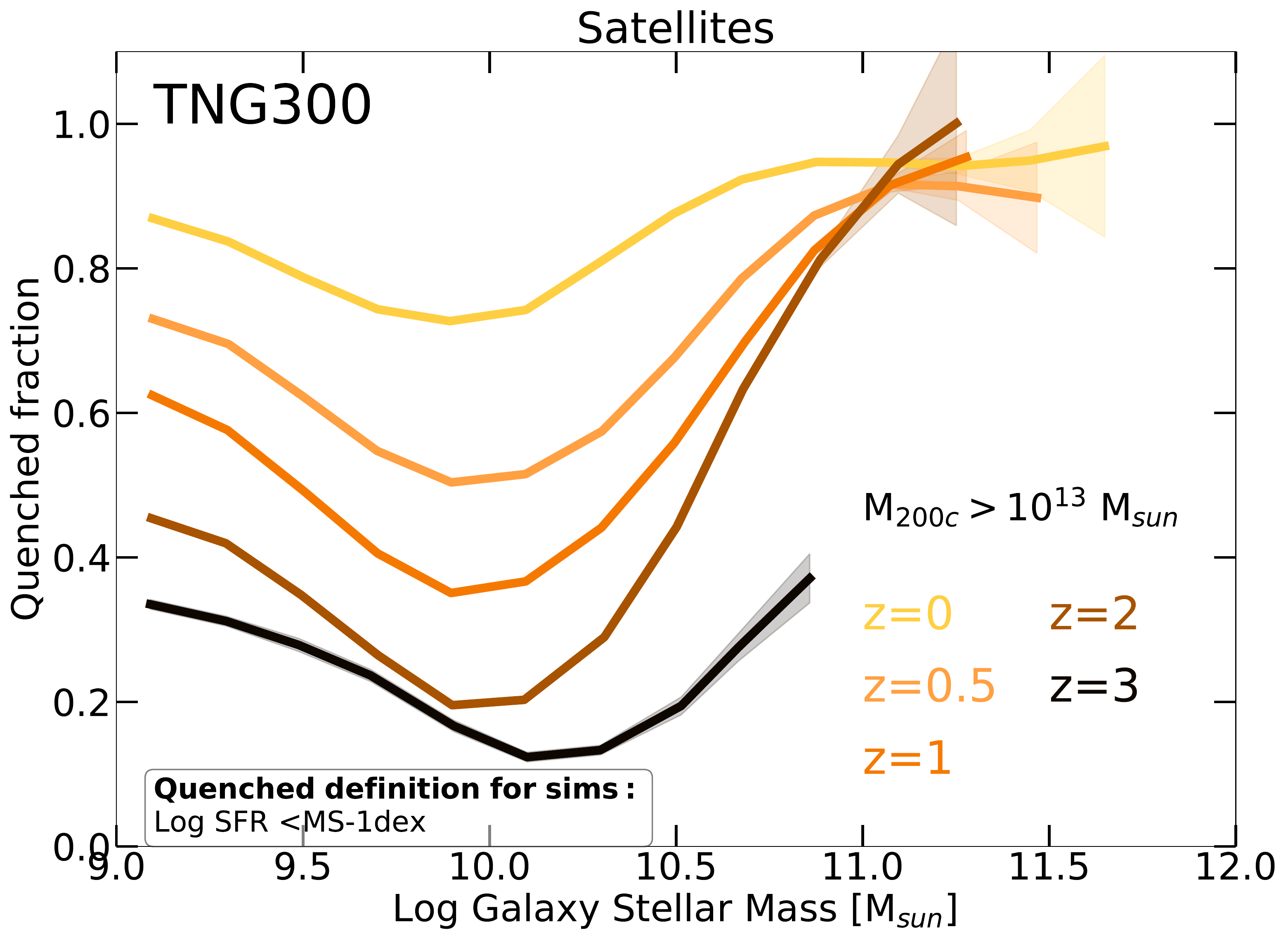}
\caption {\label{fig:frac_redshift} {\bf Trends with redshift}.
Quenched fraction versus galaxy stellar mass for TNG300 centrals (right panel) and satellites (left panel) at five different redshifts: $z=0,0.5,1,2,$ and $3$. The definition adopted to separate quenched versus star forming galaxies is Log SFR $<$MS -1dex. Shaded areas in both panels indicate the Poissonian error. For both centrals and satellites the quenched fraction is lower at higher redshift.}
\end{figure*}

A first comparison between the two panels reveals a similar percentage of quenched galaxies more massive than $\sim 10^{10.5} ~ \Ms$, about 70-80 per cent in TNG (on average), whether they are satellites or centrals. This result is consistent with the mass-quenching scenario discussed in Section \ref{intro}. In the TNG model, the halting of star formation and the reddening of galaxy colors in massive (isolated) galaxies is due to, or at least mediated by, AGN feedback and particularly by black hole driven winds at low-accretion rates \citep[see][]{2017Weinberger,2018Nelson,2020Terrazas}, which are also responsible for setting the physical and thermodynamical properties of gas within and around galaxies \citep[see][]{2020Truong,2020Davies,2020Zinger}. Within the framework of the TNG model, populations of massive central galaxies would {\it not} quench or redden at the levels expected by observational constraints (or at all) if the black hole driven winds at low-accretion rates were not in place \citep[see e.g.][]{2017Weinberger}. In other words, within the TNG model, there are no secular or internal, population-wide quenching mechanisms other than AGN feedback.

Conversely, centrals at the low mass end ($\lesssim 10^{10} ~ \Ms$) are rarely quenched, in contrast with group and cluster satellites of the same stellar mass: in TNG, the quenched fractions at the low-mass end read less than $\sim$ 5-8 per cent\footnote{ We remind the reader that some of those low-mass centrals could be backsplash galaxies.} for centrals against about 80 per cent for satellites in groups and clusters, thus pointing to the fundamental role played by a dense environment in galaxy quenching in this mass range.

These results are qualitatively consistent with a series of observational findings. 
We refer the reader to the companion paper by \cite{Donnari2020b} for a detailed comparison between the TNG results and observations at low redshift. There we show that the TNG quenched fractions at low redshift ($z\lesssim 0.6$) are in the ball park of a number of observational estimates for both centrals and satellites (their Fig. 8). More specifically, by comparing to SDSS data at $z=0.1$ and by creating detailed mocks via matching a number of analysis choices as in \cite{2012Wetzel,2013Wetzel}, we demonstrate that the quenched fractions from TNG are in striking quantitative agreement with \cite{2012Wetzel} for both centrals and satellites in the stellar mass range $10^{9.5-11.5} ~ \Ms$, when the latter are taken together for hosts with total mass between $10^{12-15}~\Ms$ (their Fig. 9, top row).

The left panel of Fig. \ref{fig:Q_frac} shows a slight difference of about 10 percentage points between TNG100 and TNG300 at both low and high stellar masses. This offset is due to a combination of statistical sampling variance owing to the significant differences simulation box sizes as well as the impact of numerical resolution. We comment on these points in more detail in Appendix \ref{appendix}. As a result of this offset, we attribute a systematic uncertainty of about 10 percentage points in the interpretation of the quenched fractions from the TNG100/TNG300 simulations. However, within the TNG framework, an even higher numerical resolution and poorer sampling statistics can impact these quenched fraction predictions with systematic uncertainties by up to 10-40 percentage points, depending on galaxy mass, host mass, and redshift \citep[see Appendix \ref{appendix} and ][for a discussion]{Donnari2020b}.

Interestingly, while the quenched fractions of centrals rise steeply from 5-8 to 80-90 per cent across the investigated stellar mass range, the quenched fractions of satellites are always as high as 80-90 per cent, although with a non-monotonic trend. We relate the slight depression at intermediate mass (more pronounced in the original Illustris, red curves) to a combination of both external (environmental) and internal (mass) quenching processes. Indeed, for low-mass satellites, the hosts in which they reside are likely the culprit for their quenching. However, at progressively higher stellar masses, the role of the environment, although still fundamental, seems to be less efficient, and as soon as the stellar mass reaches and exceeds a few $10^{10} \, \Ms$, the quenched fractions rise again, mainly driven by AGN feedback (at least in the TNG model where no other internal quenching mechanisms are in place).

The orange dot-dashed curve in the right panel represents the fraction of satellites quenched prior to their infall into any host, i.e. before becoming satellites. For galaxies less massive than $\sim 10^{10} \, \Ms$, this fraction is negligible, thus suggesting that they were mostly still star-forming when they were centrals. On the other hand, a non-negligible fraction of satellites  that today are more massive than $10^{10.5} \, \Ms$, have fallen into any host already quenched. In this case, the quenched fraction rises up to $\sim$ 50 per cent. 
In Section \ref{sec:preprocessing}, we discuss in more detail the role of pre-processing and internal processes at early times in shaping the $z=0$ quenched fraction of satellites.

While the qualitative trend with stellar mass is similar between the TNG and Illustris models, we note a quantitative difference between the two. As already highlighted in \citealt{2019Donnari}, Illustris returns quenched fractions at fixed galaxy mass that are lower than TNG by up to 0.6 for centrals and 0.3 for satellites at the transitional mass regime of around $10^{10-11} \, \Ms$. For central or high-mass galaxies, this difference could be interpreted as a shift towards higher galaxy masses for the onset of the quenching mechanism and is connected to the new and more effective AGN feedback recipes implemented in the TNG model \citep{2017Weinberger,2018Pillepich,2019Donnari}. For satellite galaxies, we speculate the difference to be due to two concurrent factors: firstly, low-mass Illustris galaxies are generally richer in both halo \citep{2018Pillepich} and HI/cold \citep{2019Diemer} gas than TNG galaxies because of the different implementation of stellar feedback; secondly, Illustris group-mass hosts are gas-poor compared to TNG and observations because of the strongly ejective AGN feedback implemented in the former \citep{2018Pillepich,2020Terrazas}. These can arguably make Illustris satellite galaxies more resilient to gas loss and Illustris hosts less effective at ram-pressure stripping.

\begin{figure*}
\centering
\includegraphics[width=0.49\textwidth]{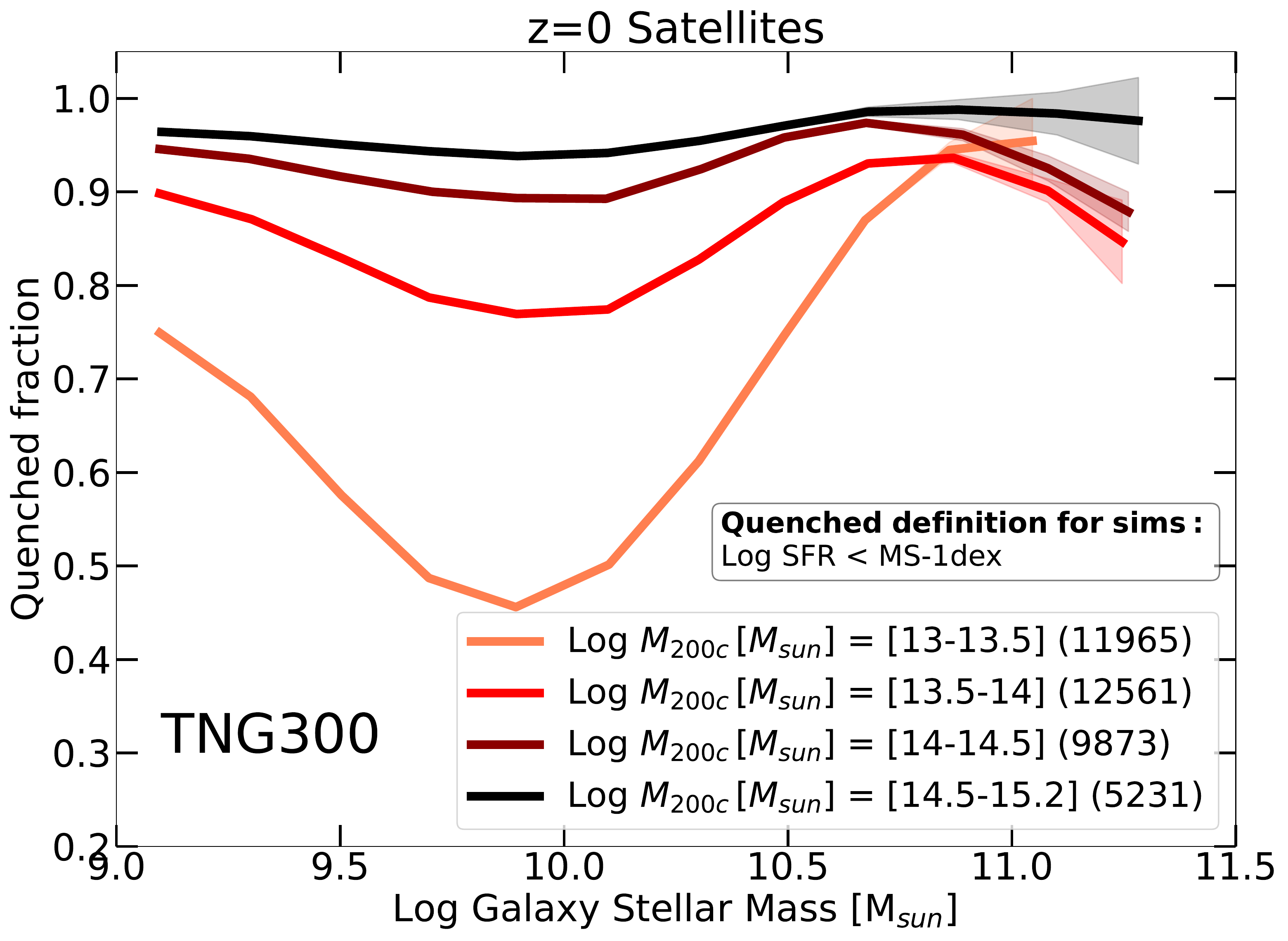}
\includegraphics[width=0.49\textwidth]{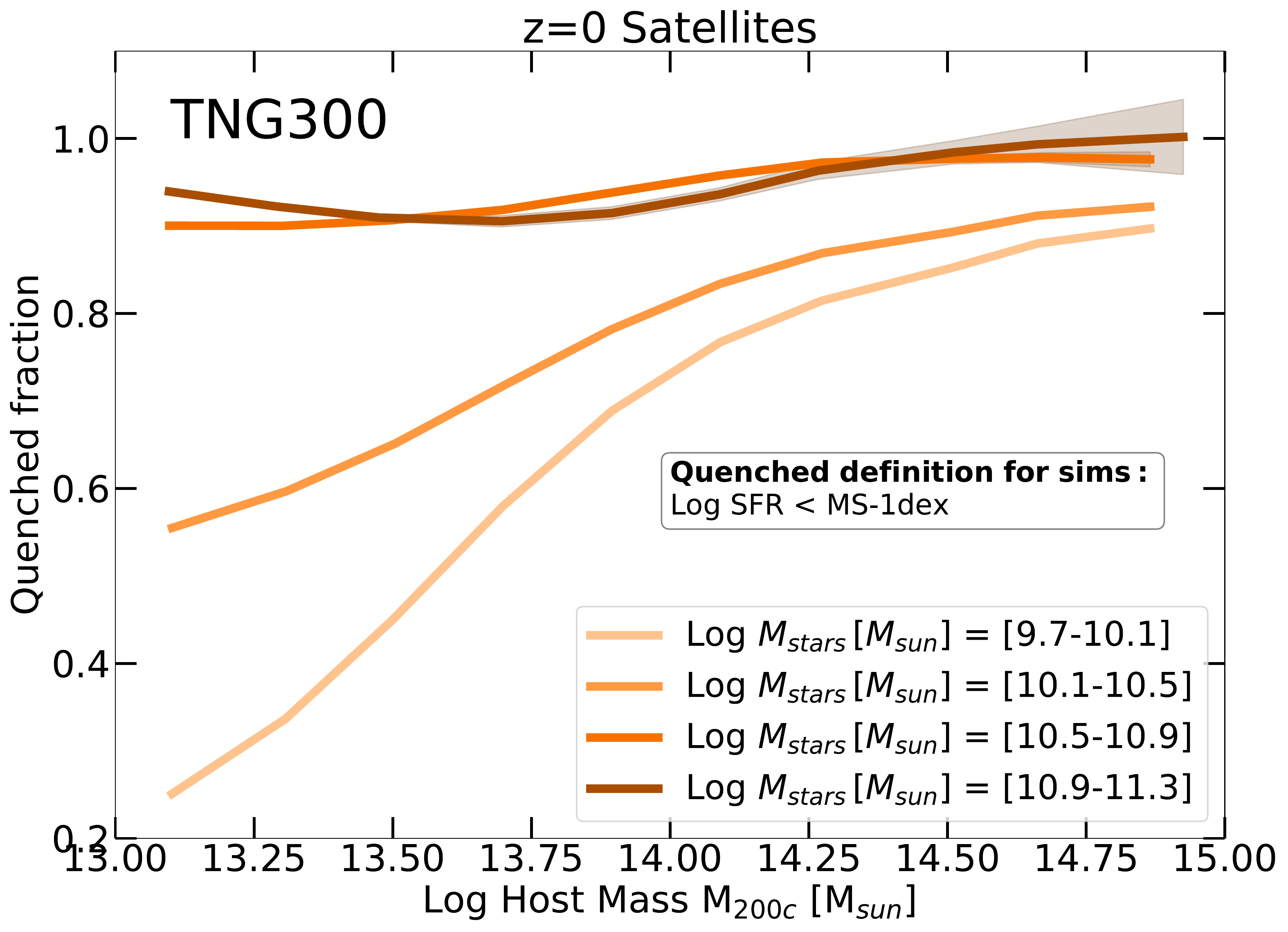}
\caption{\label{fig:environment}  {\bf Trends with stellar mass and host mass}. Left: quenched fraction as a function of galaxy stellar mass for TNG300 satellites, stacked in four host mass bins, as indicated in the legend. Right: quenched fraction as a function of $z=0$ host mass, stacked in four stellar mass bins, as indicated in the legend.
In both panels, the definition adopted to separate quenched versus star forming galaxies is Log SFR $<$ MS - 1dex and shaded areas indicate the Poissonian error. The quenched fraction of satellites is higher in massive clusters than in groups and for massive galaxies compared to the low-mass ones.}
\end{figure*}

\subsection{Quenched fractions across cosmic time}

The evolution of the SFR-stellar mass plane and the increase of the total quiescent fractions with cosmic time are well established observational \citep[see e.g.][]{2014Speagle,2014Whitaker,2018Fang} and theoretical results \citep[e.g.][for the TNG and Illustris simulations]{2019Donnari}.

In Fig. \ref{fig:frac_redshift} we show the quenched fractions of TNG300 centrals (left panel) and satellites (right panel), across five different redshifts: $z=0,0.5,1,2$, and $3$ (from yellow curves, same as Fig. \ref{fig:Q_frac}, to black curves). TNG100 (not shown) returns very similar results. We note here that the MS used to separate star-forming and quiescent galaxies is evaluated at each redshift, using the procedure described in Section \ref{QvsSF} and always extrapolated for stellar masses larger than $\MS = 10^{10.2} \, \Ms$. Moreover, as demonstrated in Figures 1-3 in \cite{Donnari2020b}, we warn the reader that different quenched definitions, SFR average timescales, and/or aperture choices can impact the quenched fraction measurements at $z=2-3$ by up to 70 percentage points.

Within the choices elected in this analysis, as expected, quenched fractions in TNG are generally lower at higher redshifts than at lower ones: while the trends with stellar mass are similar across redshifts, the fraction of quenched galaxies increases by up to 80 percentage points between $z=3$ and $z=0$ at fixed stellar mass, in both the central and satellite populations below $\sim 10^{11} ~\Ms$ \citep[][]{2016darvish,2017Fossati}. A strong redshift evolution is in place even when a narrow range of host mass is considered for the satellites: even if not shown, we have verified that for satellites residing in hosts of $10^{13-13.5}\Ms$ at any redshift, the quenched fractions is lower by up to 40 percentage points at $z=3$ than at $z=0$.
Moreover we note that the steep rise in the quenched fraction for central galaxies occurs at lower stellar masses with decreasing redshift.
On the other hand, at the very high mass end ($\gtrsim 10^{11} ~\Ms$), the quenched fraction is about 80-90 per cent regardless of the redshift.

Importantly, in TNG, a significant fraction ($\gtrsim40-50$ per cent) of galaxies is already quenched at $z\sim 2$, and more interestingly $\sim 30-40$ per cent is quenched already at $z=3$, both among centrals ($\sim10^{10-11} ~\Ms$) and satellites, suggesting that both AGN feedback and environmental processes are already in action when the Universe was a few billion years old.
This is, at least at face value, in remarkably better agreement than in previous models with the recent findings of ever more examples of very massive quiescent galaxies at $z\gtrsim 3$ (see e.g. the observational findings by \citealt{2017Glazebrook,2020Forrest} and Fig. 10 of \citealt{Donnari2020b}).


\subsection{Quenched fractions across host masses}
\label{Sec:Qfrac_acrossTime}

In this Section, we proceed to investigate how the quenched fractions vary with environment; specifically, we focus on how the quenched fractions of satellites change across host masses.

Fig. \ref{fig:environment} shows the quenched fraction of TNG300 satellites at $z=0$ versus galaxy stellar mass (left panel) and versus host mass (right panel), in bins of host and galaxy stellar mass, respectively.

We find that the non-monotonic trend of quenched fraction with stellar mass for satellites seen in the right panel of Fig. \ref{fig:Q_frac} is more  pronounced in groups (orange and red curves in the left panel) than in clusters (brown and black curves), for which the quenched fraction is almost constant at 95-100 per cent, at any stellar mass.
Moreover, as shown in the right panel, TNG returns a strongly rising trend in quenched fraction with host mass for low-mass satellites of $\MS \sim 10^{9.5-10.5} \, \Ms$, while -- importantly -- for the most massive satellites ($\MS \gtrsim 10^{10.5} \, \Ms$) the quenched fraction is 90-100 per cent at any host mass \citep[see e.g.][for similar results]{2019Davies}. 

A combination of multiple physical mechanisms can be seen in these plots. Firstly, the independence on host mass of the quenched fractions for massive galaxies (orange and brown curves in the right panel of Fig.~\ref{fig:environment}) suggests that either more massive satellites are less susceptible to environmental processes or are likely to be quenched mostly due to their internal secular processes, or a combination thereof. The fact that, at least in TNG, the quenched fractions of massive galaxies ($\MS \gtrsim 10^{10.5-11} \, \Ms$ ) are beyond 70-80 per cent whether they are centrals or satellites (see also Fig.~\ref{fig:Q_frac}) strongly points to internal quenching processes being the main actors also in high-density environments: in the TNG model, the only internal mechanism that is effective at quenching is feedback from super massive black holes, as already discussed in Section \ref{sec:qfrac}. Secondly, more massive hosts are likely to impart stronger satellite quenching effects, thus quenching their galaxies more efficiently. 
However, as we will discuss in more detail in Section \ref{sec:preprocessing}, one must take into account that a satellite might be orbiting in other groups before being accreted into the current one. As a result of this group pre-processing, these satellites might fall into the $z=0$ host already quenched. 

All the results of Fig. \ref{fig:environment} are qualitatively consistent with SDSS-based results compiled by \cite{2012Wetzel, 2013Wetzel}, who focus on galaxies more massive than $10^{9.7} \Ms$. In our companion paper (\cite{Donnari2020b}) we demonstrate that the agreement with SDSS in quenched fractions is best (and really good, within 5 percentage points) at intermediate host masses ($10^{13-14} \, \Ms$) and galaxy stellar masses ($10^{10-11} \, \Ms$), while the TNG quenched fractions of $10^{9.5-10} \, \Ms$ satellites in more masive hosts ($10^{14.5-15} \, \Ms$) are higher by up to 20 percentage points than those inferred from SDSS.

\begin{figure*}
\centering
\includegraphics[width=12cm]{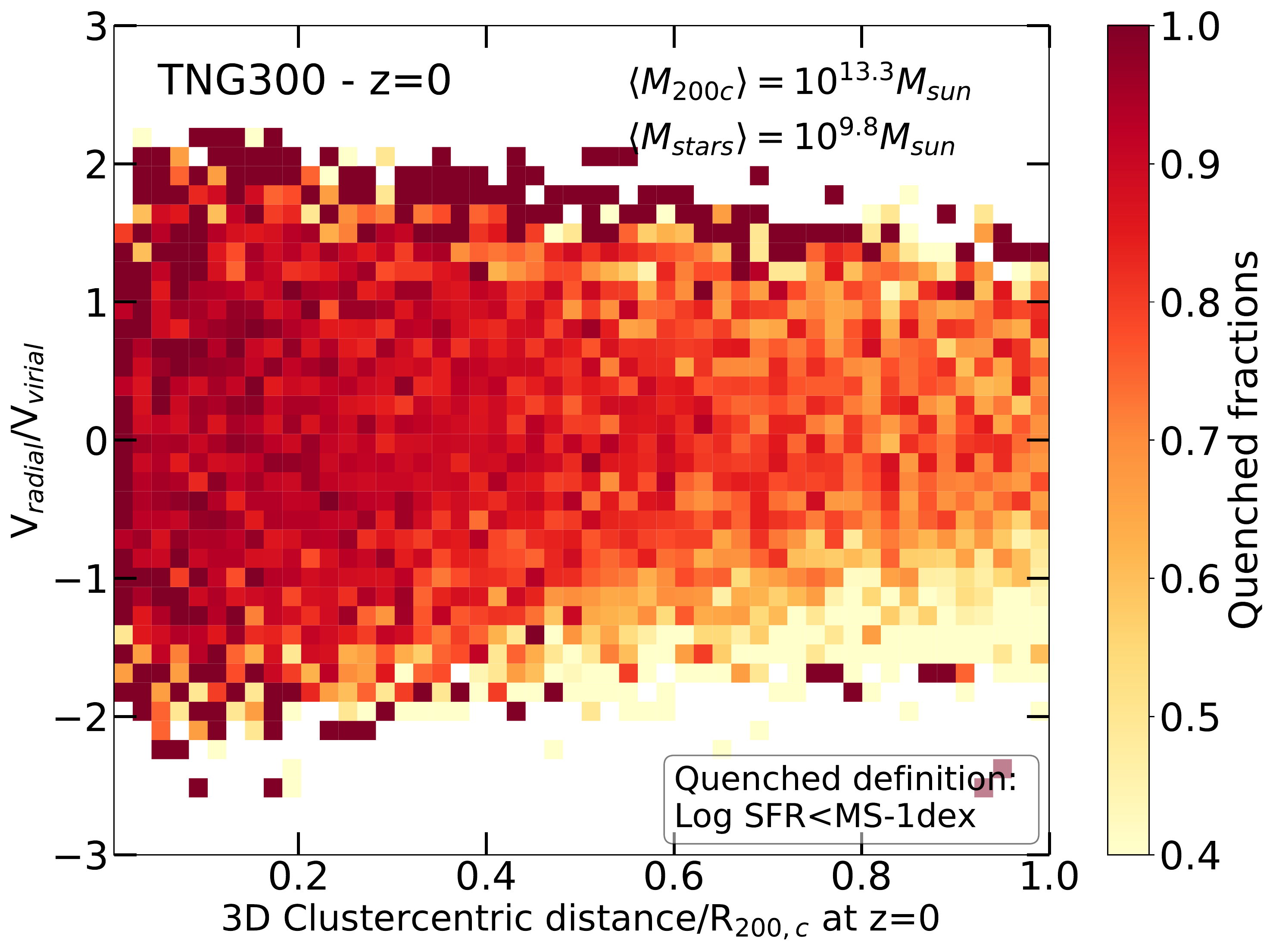}
\includegraphics[width=0.46\textwidth]{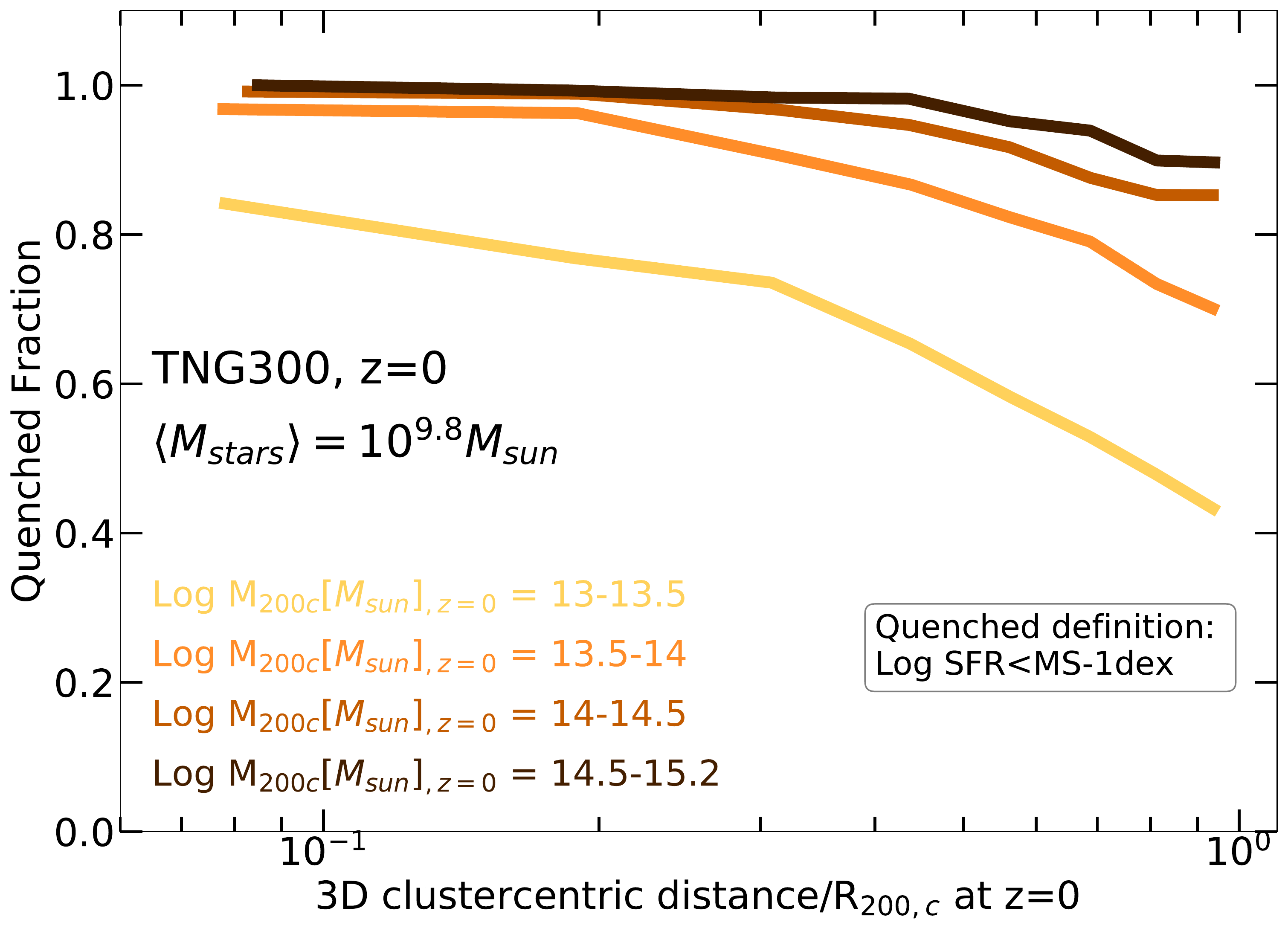}
\includegraphics[width=0.46\textwidth]{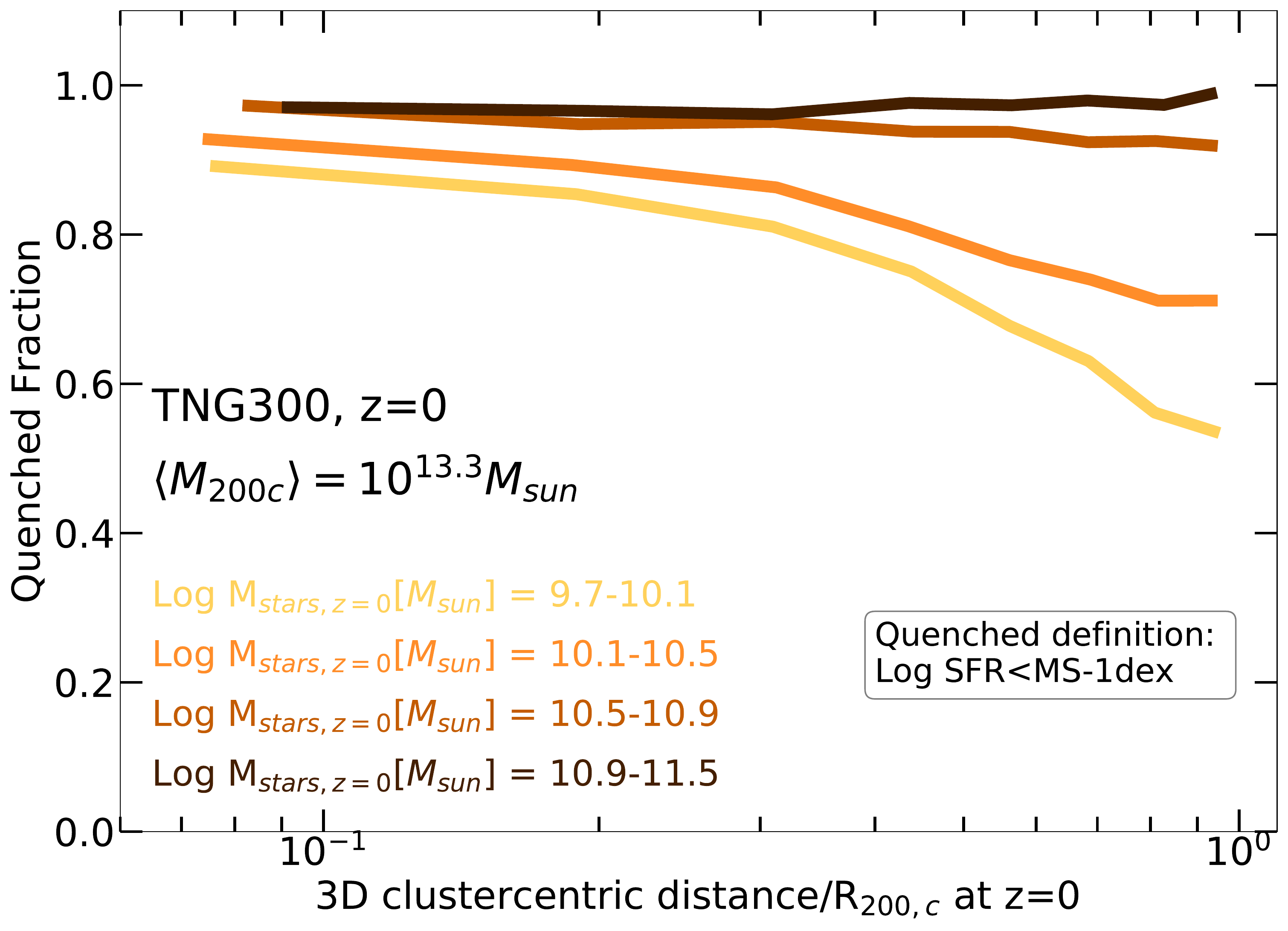}
\includegraphics[width=0.46\textwidth]{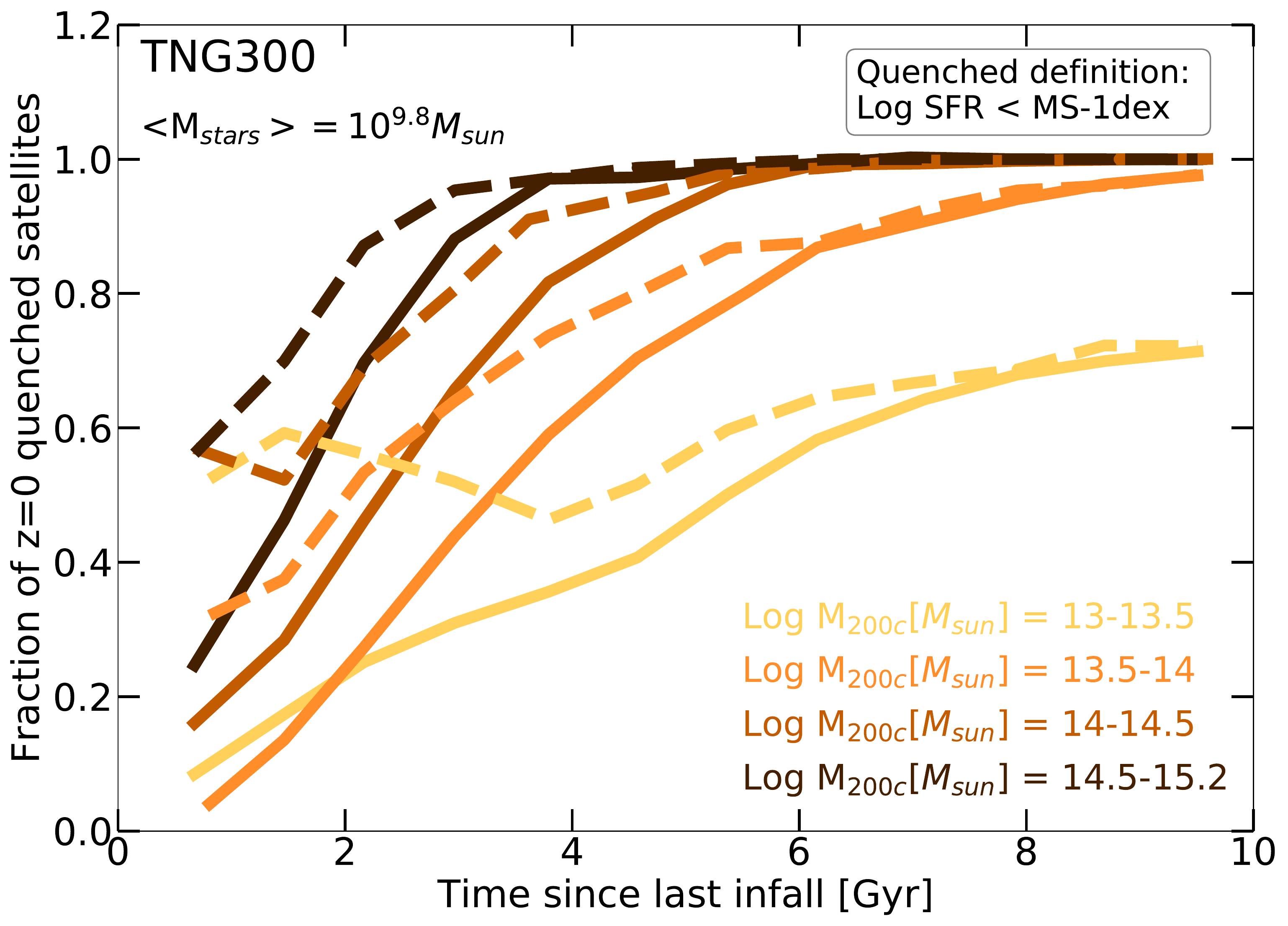}
\includegraphics[width=0.46\textwidth]{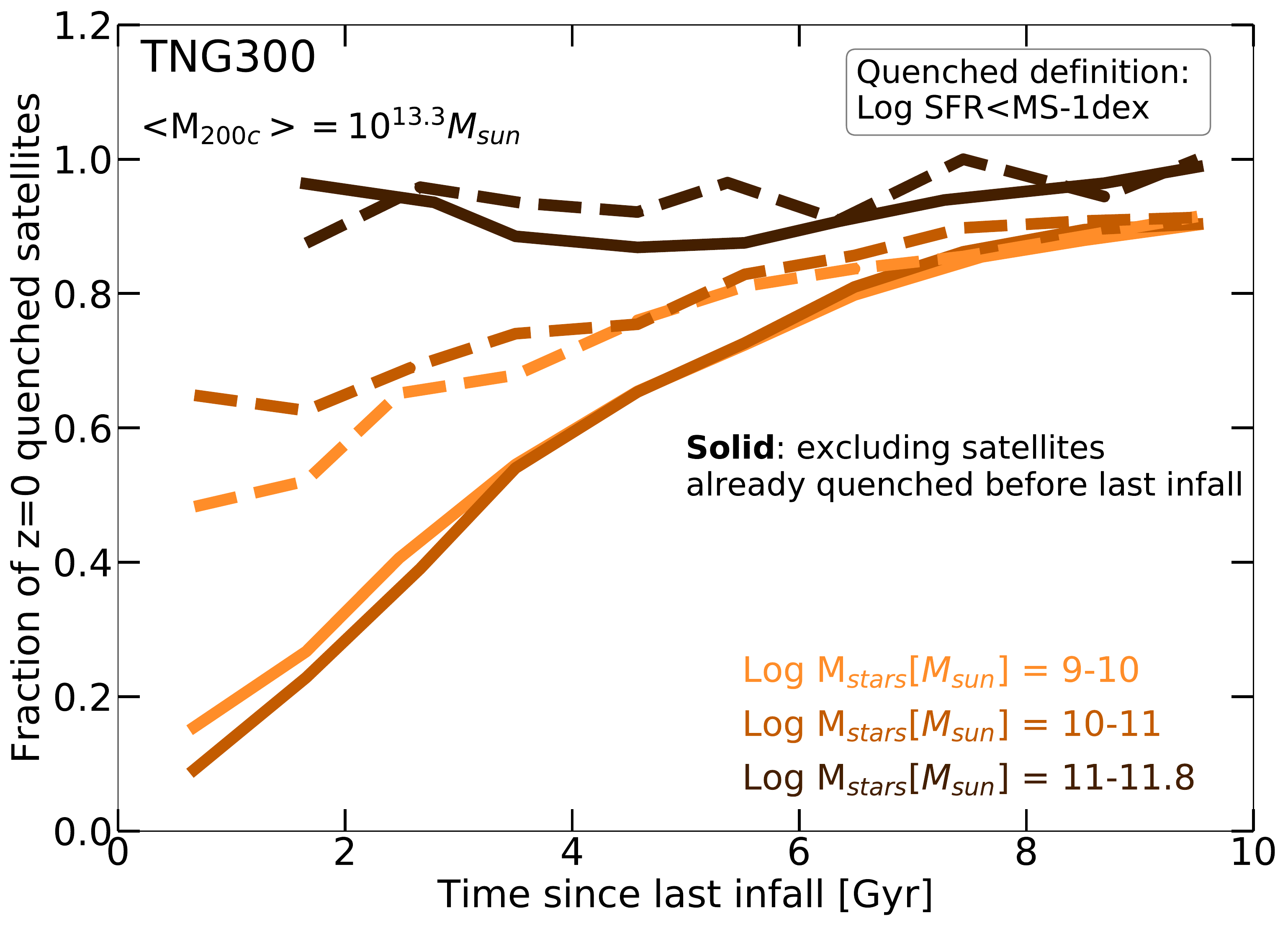}
\caption {\label{fig:frac_r2} {\bf Dependence on cluster-centric distance and time since last infall.} Top: phase-space diagram for TNG300 satellites ($\MS\geq 10^9 ~\Ms$) in a range of different hosts ($M_{200c}\geq 10^{13} ~\Ms$) at $z=0$ where the color indicates the median of the quenched fractions in each pixel of the grid.
Middle row: fraction of quenched galaxies in TNG300 at $z=0$ as a function of their 3D distance from the cluster center, scaled to the host virial radius $R_{200c}$ with galaxies stacked according to their present-day halo mass $M_{200c}$ (left panel) and their present-day satellite mass (right panel). 
Bottom: Fraction of quenched galaxies (at $z=0$) in TNG300 versus the time since their last infall, stacked in four host mass bins (left panel) and three stellar mass bins (right panel). In both panels, masses are measured at $z=0$ and the quenched fraction is evaluated including (dashed curves) or excluding (solid curves) satellites already quenched at the time of their last infall.
In all plots, the quenched definition is Log SFR$<$MS-1 dex. 
The fraction of passive galaxies is higher closer to the host center. This is not the case for massive satellites, for which the quenched fractions are $\sim$90 per cent irrespective of their distance from the host. Early infallers are more likely to be quenched compared to recent infallers.}
\end{figure*}

\subsection{Satellite phase-space diagram, $z=0$ radial distribution of satellites within $R_{200c}$ and connection to the time since last infall}
\label{sec:PhaseSpace}
Observations suggest that the satellite quenched fractions also depend on location within a host (see Introduction).
The top panel of Fig. \ref{fig:frac_r2} shows the median fraction of quenched galaxies in TNG300 at $z=0$ in the plane of satellite radial velocities (normalized to the host virial velocity measured as $\sqrt{G ~(M_{200c}/R_{200c})}$ versus their 3D distance to the cluster center (normalized by the virial radius $R_{200c}$). Averages are computed in 2D bins of size 0.1 dex in distance and 0.3 dex in velocity space.

A clear trend is manifest at any fixed value of $V_{\rm radial}/V_{\rm virial}$: the median quenched fraction decreases at increasing distance to the host center. 
As we demonstrate later when discussing the middle panels, this trend is mostly due to low-mass satellites, while the fraction of the massive satellites is about 90 per cent regardless of the distance from the host centre.
Additionally, at a fixed small distance from the cluster center ($\lesssim0.6 R_{200c}$), infalling ($V_{\rm radial}<0$) and outgoing ($V_{\rm radial}>0$) satellites exhibit nearly identical quenched fractions. On the other hand, infalling satellites residing at a distance larger than $\sim 0.6 R_{200c}$ show a quenched fraction lower than the outgoing satellites, at the same cluster-centric distance.
We speculate that satellites that have already orbited close to the host center and have spent more time in the hosts -- as those in outbound orbits and positive radial velocity -- are more likely to be quenched compared to those that are still infalling or on their first infalling trajectory.

An analogue trend, in overall qualitative agreement with our results, is described in \cite{2019Pasquali}, where the authors combine SDSS group catalogues and the YZiCS hydrodynamical simulations  \citep{2017Choi}. Their results reveal that the passive fraction increases -- from 50 per cent to more than 90 per cent -- at decreasing projected cluster-centric distance.

In the middle panels of Fig.~\ref{fig:frac_r2} we therefore show quenched fractions versus 3D distance by stacking satellites in four host mass bins (right panel) and four stellar mass bins (left panel).
In general, we find that the fraction of low-mass passive galaxies is higher closer to the host center, consistent with the top panel results just discussed. However, the trend with clustercentric distance is stronger the lower the host mass is. Importantly, this trend is very weak, if not nonexistent, in the case of massive satellites ($\gtrsim10^{10.5} ~ \Ms$) and satellites in massive hosts ($\gtrsim10^{14} ~ \Ms$), for which the quenched fractions are larger than 80-90 per cent irrespective of distance from the host center. 
Overall, as shown in \cite{Donnari2020b}, the trend with halocentric distance is well consistent with SDSS estimates \citep{2012Wetzel}, with the level of (dis)agreement depending on host and galaxy stellar mass ranges, being best (and in fact, excellent) for intermediate hosts of $10^{13-14}\Ms$ and worst for the lowest mass satellites in either the lowest ($\sim 10^{12}\Ms$ ) or highest mass hosts ($10^{14.5-15}\Ms$).

This confirms, once again, the fundamental role played by the environment in quenching low-mass satellites, while the quenching of massive ones is almost independent of the host in which they reside and hence, arguably mainly related to their internal processes.

A satellite's location within a host is correlated with the time since accretion: satellites that fell into their current hosts a longer time ago are typically found closer to their host centers \citep{2017Rhee}. Therefore the trends of the top panels of Fig.~\ref{fig:frac_r2} can be attributed to two concurrent facts: smaller clustercentric distances correspond to denser regions where environmental effects can be larger but also, even though individual orbits of satellites can span a broad range of radii, satellites orbiting closer to the cluster center have typically spent a longer time within their group or cluster environments: if the fraction of quenched satellites depends on how long they have been orbiting in their $z=0$ host \citep[a plausible assumption, see also][for similar results]{2020Pengfei}, then the probability of being quenched can be higher closer to the host centers than in the outskirts, irrespective of environmental density.

In the bottom panels of Fig. \ref{fig:frac_r2} we show the fraction of quenched satellites as a function of the time since their last infall into the current host (see Section \ref{sec:preprocessing} for a detailed definition of infall times), by binning galaxies by their present-day host mass (left panel) and their present-day stellar mass (right panel).
In both panels, the fraction of galaxies is evaluated including (dashed curves) or excluding (solid curves) satellites that were already quenched at the time of their accretion onto the current host -- we postpone to Section \ref{sec:preprocessing} for a more detailed analysis on pre-processing.

A clear result is evident: the fraction of $z=0$ quenched satellites is higher for those accreted earlier into the current host, this correlation being stronger if we remove galaxies already quenched at infall (solid versus dashed curves) and for more massive hosts and lower-mass satellites. 
Interestingly, the trend flattens completely at 80-100 per cent for satellites that were accreted longer than 4-6 Gyr ago in hosts more massive than $10^{13.5} \Ms$: this suggests an upper limit of some billion years for the time needed for the group and cluster environment to have an effect and to induce quenching. Allowing for some delay, this estimate is in line with the results of \citealt{2019Yun}, where we found that jellyfish galaxies (i.e. galaxies with markedly visible indications of ram-pressure stripping) are recent infallers according to the TNG simulations, i.e. they fell no more than 2-3 billion years before being identified as jellyfish galaxies. 

\section {Disentangling quenching pathways}
\label{discussion}

In Section~\ref{results}, we have shown that in TNG, massive galaxies ($\MS \gtrsim$ a few $10^{10}~ \Ms$) display high quenched fractions, regardless of whether they are centrals or satellites and in the latter case, regardless of host mass, cosmic time, cluster-centric distance and time since infall. 
This would suggest that the massive satellites observed to be quenched at $z=0$ could have halted their star formation before ever becoming satellites, or even if they were quenched as satellites, could have done so due to internal or secular quenching mechanisms\footnote{As discussed in the previous Sections, in the framework of the TNG model there are no emerging or explicitly-modeled internal or secular quenching mechanisms other than the feedback from super massive black holes. Therefore, in the following discourse, we will associate secular/internal quenching mechanisms with AGN feedback.} rather than any environmental effects.

Similarly, in the previous Section we have shown that regardless of galaxy stellar mass and cluster-centric distance, the fraction of quenched satellites orbiting in massive clusters of $M_{200c} \gtrsim 10^{14} ~ \Ms$ is higher than 80-90 per cent. 
A large body of work \citep[see e.g.][and references therein]{2017Henriques}, mostly based on analytical arguments and controlled experiments, supports the idea that would be deduced from this finding, namely that more massive clusters are more efficient than lower-mass groups at quenching their satellite galaxies. This would be because they provide higher-density environments, whether in terms of deeper potential wells for the purposes of tidal stripping and galaxy encounters, higher galaxy number density for the purposes of harassment, or higher ambient gas density for the purposes of ram-pressure stripping.

However, the hierarchical growth of structure complicates the straightforward interpretation of e.g. Figs.~\ref{fig:Q_frac} and \ref{fig:environment}, because the most massive clusters in today's Universe formed from the merging of smaller groups that in turn have hosted their own satellite populations. Furthermore, hosts of different total mass exhibit different infall time distributions of their galaxy members \citep{2020Engler} and the longer the exposure to environmental processes, the higher the probability that a galaxy quenches (Section \ref{sec:PhaseSpace}).  

In this section we therefore proceed to address the following questions: 
do high-density environments quench their satellites more efficiently with respect to lower-mass groups even after taking into account pre-processing? And what is the role of lower-mass groups in quenching galaxies prior to their accretion onto the $z=0$ host? 
Theoretical models offer a unique opportunity to shed light on the timing and circumstances whereby quenching occurs, because of their capability to follow galaxies back in time i.e. for example to follow satellite galaxies until their infall into any host. We hence use the TNG simulations to systematically address these questions in the following subsections.

\begin{figure}
\centering
\includegraphics[width=\columnwidth]{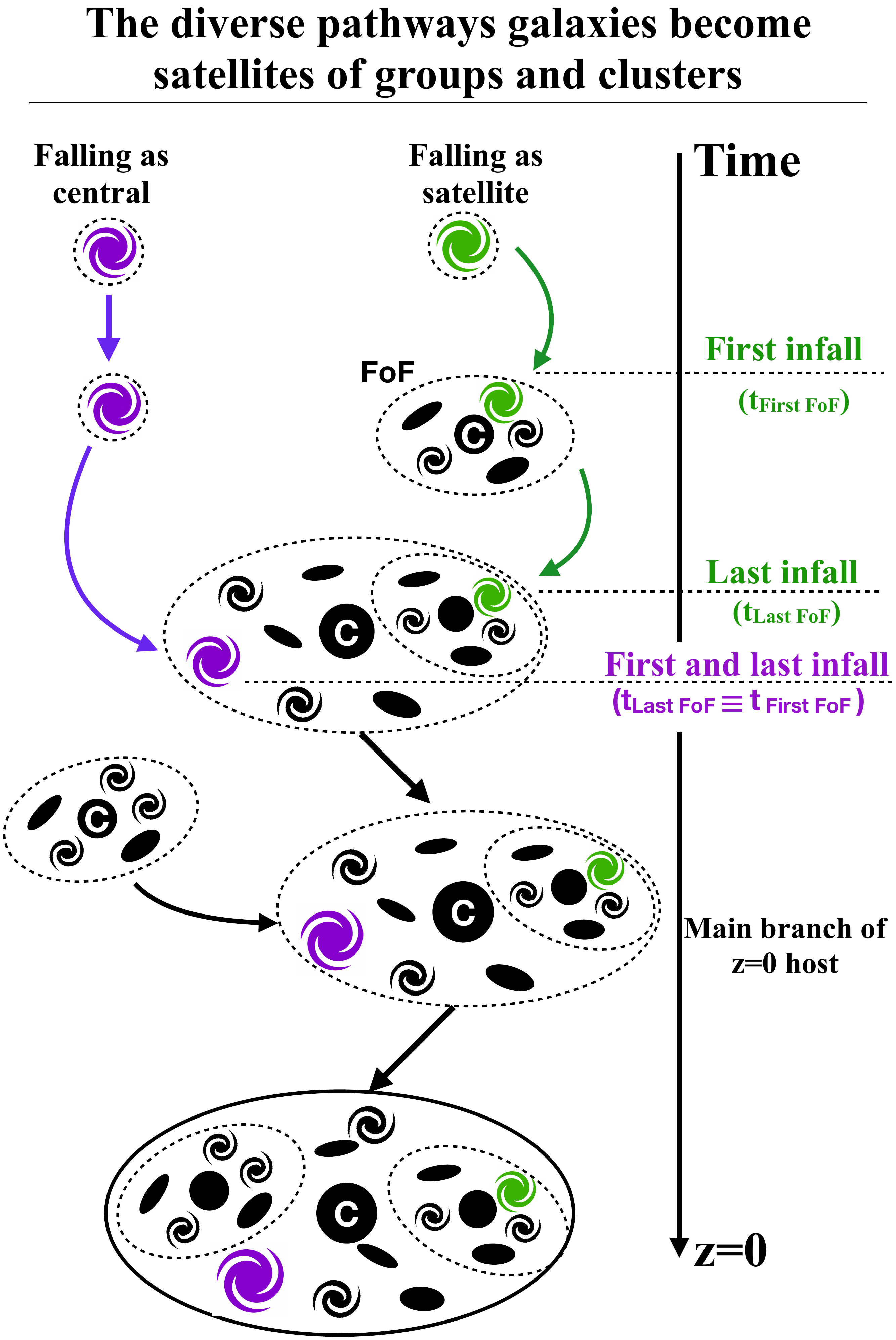}
\caption{\label{fig:sketch} {\bf The concept of \textit{pre-processing} and the evolutionary pathways of $z=0$ satellites}. The schematic cartoon describes two possible evolutionary tracks for satellites inhabiting groups or clusters at $z=0$. 
A galaxy might be directly accreted into the main branch of the $z=0$ host and thus have only one infall time (falling as centrals, purple). On the other hand, a galaxy might first fall into a subgroup rather than into the $z=0$ one, which in turn will merge in the final $z=0$ host (falling as satellite, green). We call these satellites ``pre-processed''.
Clearly, for some galaxies the situation might be more complex, as we discuss in the text.}
\end{figure}
\subsection{The concept of pre-processing}
As mentioned in Section \ref{intro} and already alluded to in the above TNG-based analyses, both observations and models recognize that environmentally-driven transformations may occur in satellites even before they were accreted into their current host.
Because of the nature of the hierarchical growth of structure, we expect the $z=0$ population of satellites residing today in groups and clusters to also include galaxies that orbited in lower-mass hosts before falling into (the progenitor of) their current host: this mechanism is commonly referred to as \textit{pre-processing}.

Particularly, in this paper we describe as \textit{pre-processed} all those satellites that have been residing in ``subgroups'' of total mass of at least $10^{12} ~\Ms$ before being accreted by the progenitor of their host at $z=0$ (or at the time of observation). The choice of a minimum pre-processing host mass is required to carry out an even-handed analysis between simulations with varied resolutions or halo identification definitions.  Here, we have employed a minimum halo mass of $10^{12} \MS$ because we are interested in identifying galaxies that may have experienced some levels of environmental processes in the past, e.g. environmental processes that may be capable of quenching galaxies.

In fact, galaxies that are found today in groups and clusters may have followed {a diverse range of} pathways to become satellites. 
For simplicity, in Fig. \ref{fig:sketch} we illustrate the two main possible scenarios. A galaxy may directly fall into the main branch of its $z=0$ host and remain there until $z=0$ (purple galaxy). We name this category {\bf falling as centrals}. 
On the other hand, a satellite may be accreted by a subgroup at $z>0$ that, due to the hierarchical growth of structure, might in turn merge into the main progenitor of a $z=0$ group or cluster. We name these satellites {\bf falling as satellites} or, indeed, {\bf pre-processed}.

Note that our definition of pre-processing so far does not yet assert anything as to whether the galaxy properties are affected by their pre-processing host. 
In fact, how much the pre-processing contributes to the present-day quenched fractions of cluster galaxies is still a matter of debate (see Section \ref{intro}) and will be addressed in Section~\ref{sec:contributions}.

Furthermore, the analysis of cosmological simulations clearly shows that the situation is generally more complex than the one depicted in Fig. \ref{fig:sketch} and other pathways are possible. For example, a galaxy, after being captured by a subgroup, might have an orbit sufficiently wide to leave the host after a few hundreds of Myr, or in some cases after a few Gyr, thus becoming a central again. It could then be accreted by a subgroup or directly onto the main branch of the final host. Alternatively, a galaxy, after being accreted by its final host, can escape again for a non-negligible amount of time and then come back into the main branch of the $z=0$ host. Even though these further situations are plausible, throughout the paper we consider those scenarios as part of the ''pre-processed`` one for simplicity, without making any distinction in our plots unless otherwise indicated.

\begin{table}
\centering
\begin{tabular}{l|c}
\hline
{\large Infall time} & {\large Description} \\
\hline
\hline

t$_{\rm First FoF}$ & \multicolumn{1}{m{6cm}}{The time when a galaxy becomes satellite of any FoF (of $M_{200c}>10^{12} \, \Ms$) for the first time.}\\
t$_{\rm Last FoF}$ & \multicolumn{1}{m{6cm}}{The time when a galaxy is accreted onto its last host, remaining there until $z=0$.}\\

\hline
\end{tabular}
\caption{\label{tab:infalltime} Description of the two infall times used in this work (see also Fig. \ref{fig:sketch} for a visual schematic).}
\end{table}
\subsubsection{Infall times for galaxies falling as satellites and as centrals}
\label{Infalltime}

 The search and labeling of pre-processed satellites is practically implemented in our analysis via the definition of two different infall times.
As a reminder, our sample of satellites is selected at $z=0$ and includes galaxies with stellar mass $\geq 10^9 \, \Ms$ residing within one virial radius $R_{200c}$ of their host with $M_{200c} \geq 10^{13} ~ \Ms$. 
The so-called backsplash galaxies, i.e. satellites found today outside the virial radius of the host where they were previously orbiting, are hence neglected. Furthermore, the time evolution of each satellite in the simulations is followed using the \texttt{SUBLINK} algorithm, which allows us to reconstruct the merger trees at the subhalo level and to study the properties of each galaxy and its host progenitor by following the main branch of the merger tree \citep{2015Vicente}.

Tracing a galaxy back in time, we can define (at least) two different infall times\footnote{In principle, the definition of infall time can be different from work to work, often invoking the crossing of the virial radius. For clarity, in this paper we only use the two defined in the text and we postpone to a future study the task of investigating in more detail the implication of different choices.}, as summarized in Tab. \ref{tab:infalltime} and depicted in the schematic of Fig. \ref{fig:sketch}: 

\begin{enumerate}
\item The time when a galaxy becomes a satellite of a FOF for the first time: we will refer to this time as $\rm t_{\rm First FoF}$. We note here that, in order to consider a galaxy as part of a (different) halo, we require a minimum of three consecutive snapshots in which it is a satellite and a minimum host mass (at the time of infall) of $10^{12} \, \Ms$.
\\
\item The time when a galaxy falls -- for the last time -- into the main branch of its current host and remains there until $z=0$: we will refer to this time as $\rm t_{\rm Last FoF}$. 
\end{enumerate}

Therefore, in the case of satellites falling as centrals, the two infall times described above coincide: $\rm t_{\rm First FoF} \equiv  \rm t_{\rm Last FoF}$. On the other hand, for galaxies falling as satellites (i.e. pre-processed) we can identify two distinct infall times, with the accretion into a first host occurring at an earlier cosmic time with respect to the accretion into the current host: $\rm t_{\rm First FoF} < \rm t_{\rm Last FoF}$.

\subsection{Demographics of pre-processed satellites in the TNG simulations}
\label{sec:preprocessing}

\begin{figure*}
\centering
\includegraphics[width=14cm]{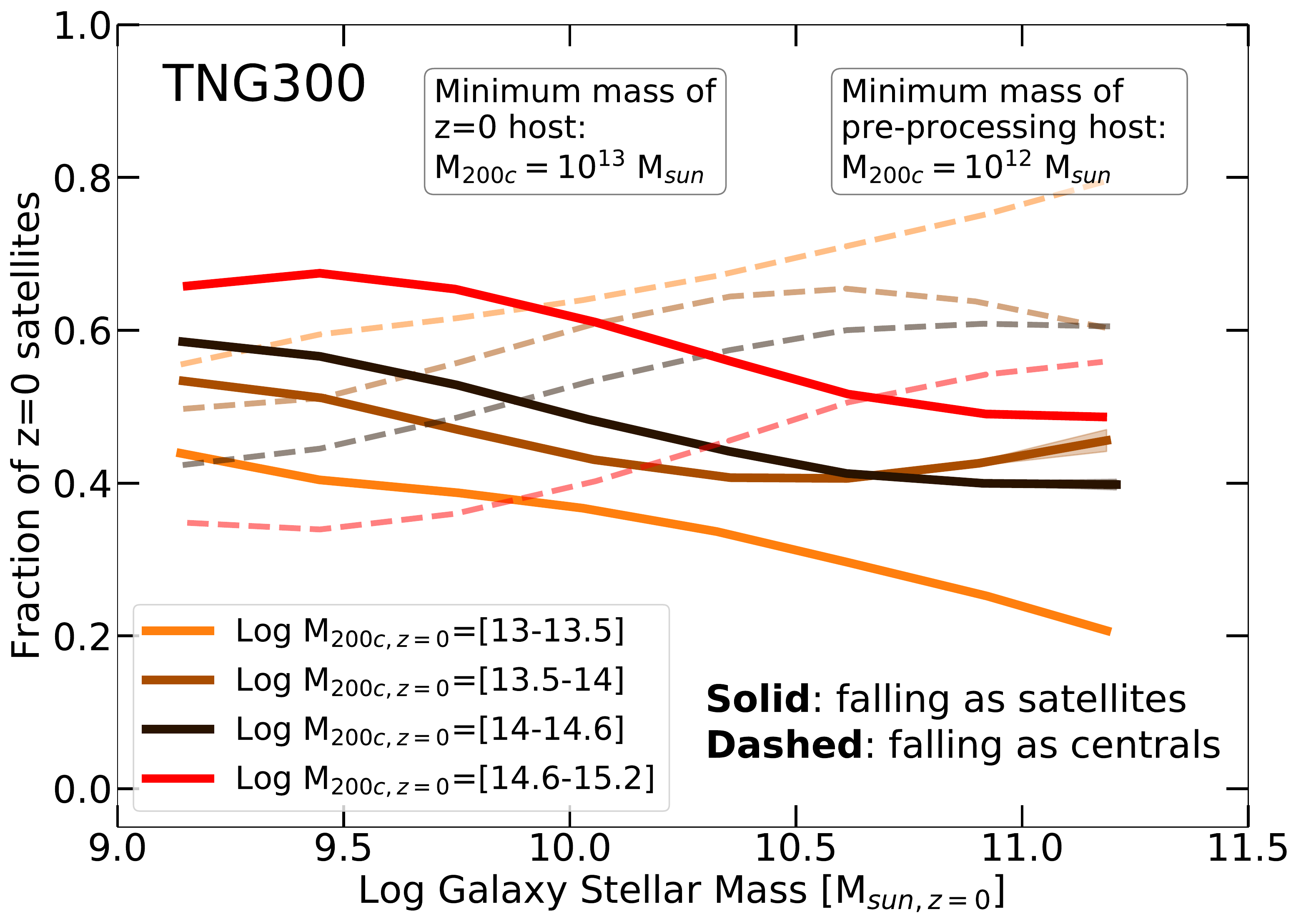}
\includegraphics[width=0.49\textwidth]{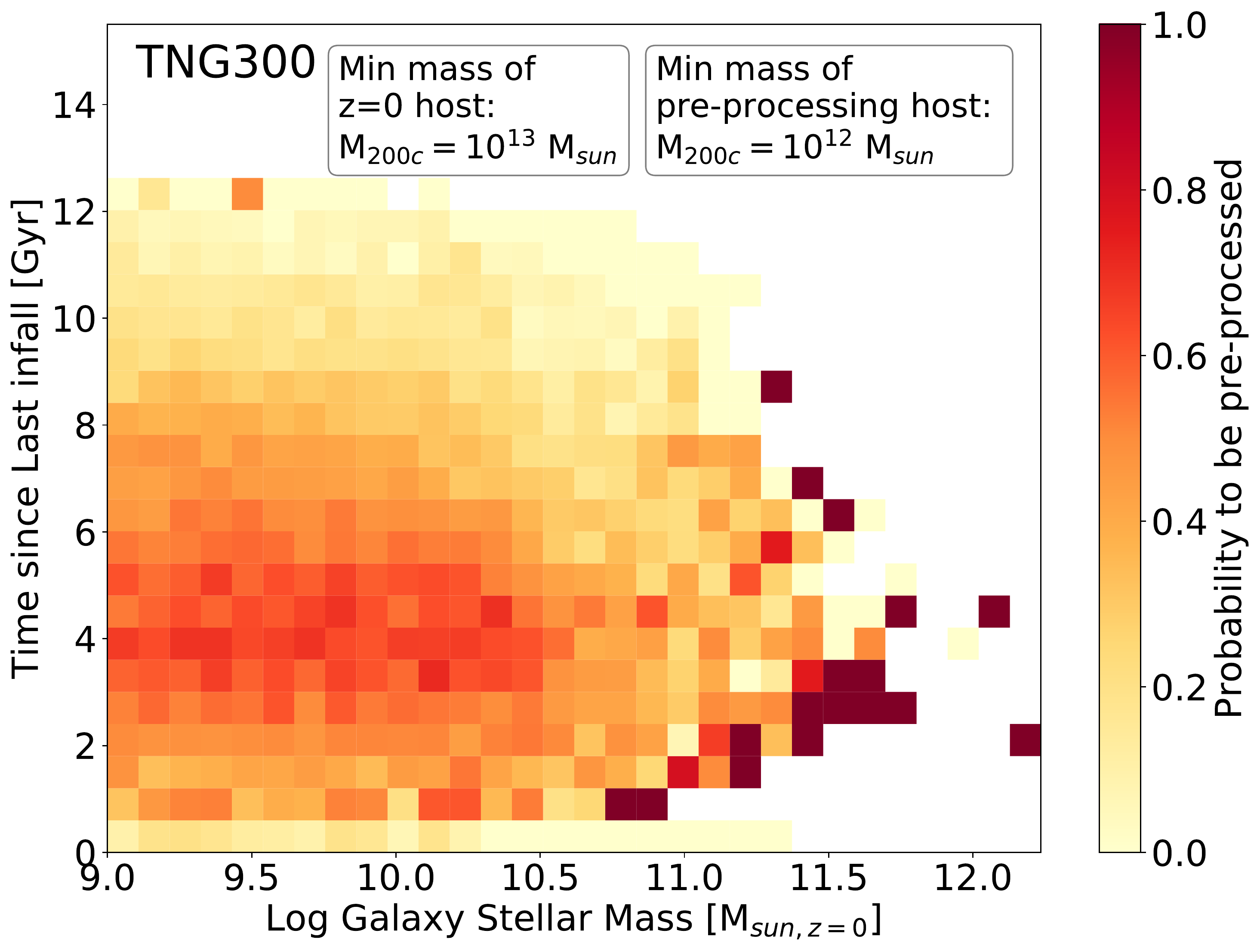}
\includegraphics[width=0.49\textwidth]{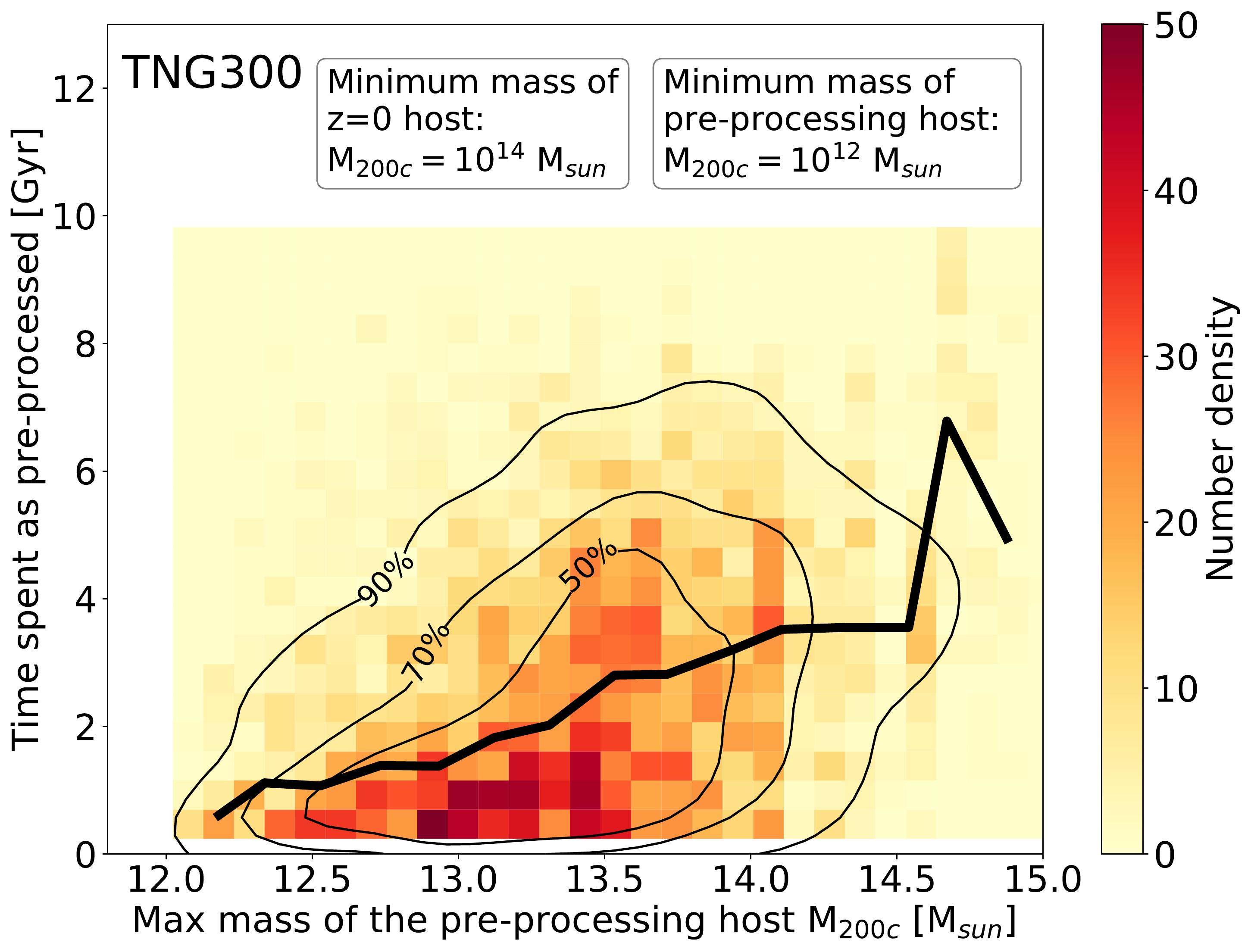}
\caption{\label{fig:qfrac_preproc} {\bf Pre-processing statistics for group and cluster satellites}. Top: fraction of TNG300 $z=0$ satellites that have been pre-processed (i.e. fallen in as satellites, solid) or have fallen in as centrals (dashed), as a function of galaxy stellar mass. The fraction of pre-processed satellites is higher in higher mass hosts at $z=0$. Bottom left: time since last infall of TNG300 satellites (into the $z=0$ host) as a function of galaxy stellar mass. The color represents the probability of having been pre-processed by other hosts with a minimum mass of $M_{200c} > 10^{12} \, \Ms$, before falling into the $z=0$ host. Bottom right: time spent as a pre-processed satellite as a function of the peak host mass. The color denotes the number density of galaxies while contours encompass 50, 70 and 90 percent of the galaxies sample. The higher the mass of the pre-processing host, the longer the time a satellite spent there.}
\end{figure*}

\begin{figure*}
\centering
\includegraphics[width=0.47\textwidth]{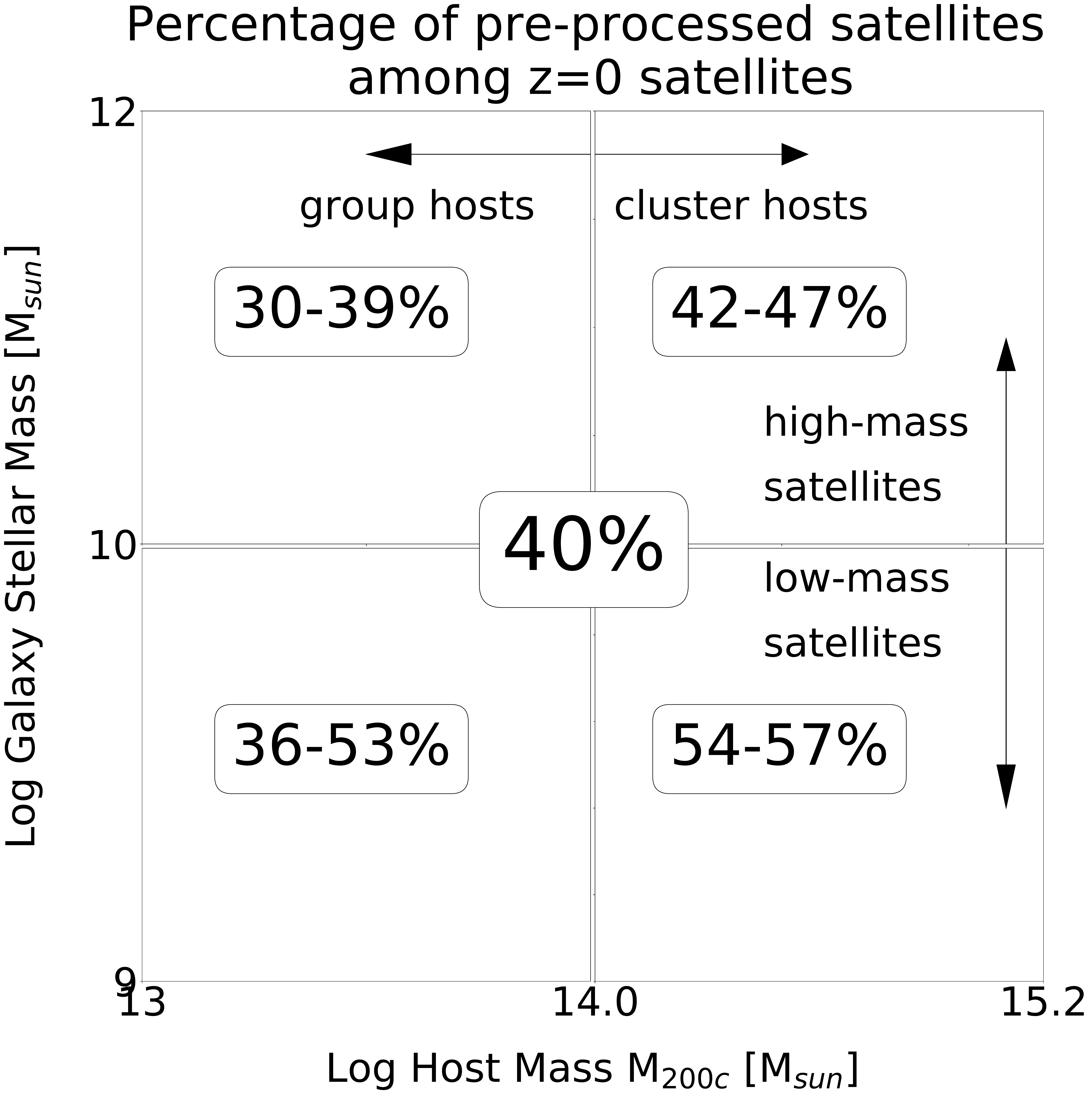}
\includegraphics[width=0.47\textwidth]{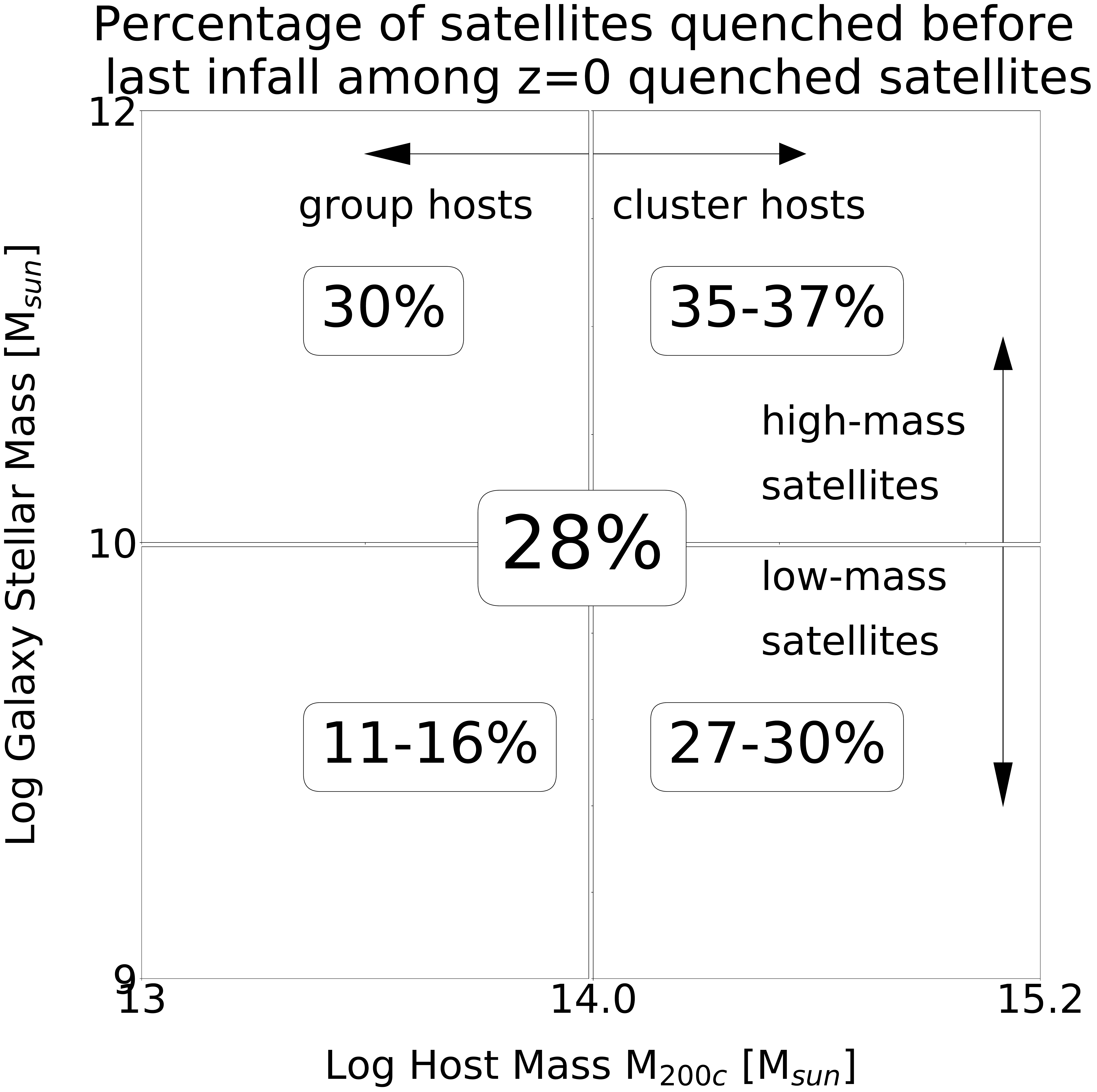}
\caption{\label{fig:preproc_prequench} {\bf Percentages of pre-processed satellites (left) and of satellites quenched before last infall (right).} The fractions given in the left represent the number of $z=0$ satellites that have been pre-processed, i.e. have orbited in different subgroup prior to infall into their $z=0$ host. The right panel depicts the fractions of satellites that quenched before falling into their current host over the $z=0$ quenched population. 
Each panel is divided into four tiles, denoting low and high stellar masses (below or above $\sim 10^{10}\,\Ms$) and residing in groups and clusters (below or above $\sim 10^{14}\,\Ms$). The numbers in each tile bracket the results from TNG100 and TNG300, respectively. Pre-processing occurs more frequently in clusters than in groups, regardless of stellar mass. Among $z=0$ quenched satellites, about 30 per cent are already quenched at the time of their last infall.}
\end{figure*}

With our definition of pre-processing and of satellites falling as centrals or as satellites in hand, we now proceed to study the demographics of these two populations in the TNG simulations. We first aim to quantify how {\it frequent} pre-processing is, based on the histories and orbits of TNG satellites and hosts, and with no regard for the internal properties of galaxies. In the next Section, we will quantify how {\it effective} pre-processing is at determining the $z=0$ quenched fractions.


In the top panel of Fig. \ref{fig:qfrac_preproc}, we show the fraction of $z=0$ TNG300\footnote{Although not shown, the same qualitative trends with both stellar and host mass also hold in TNG100. However, we note that due to a combination of different sampling of differently-sized volumes (chiefly) and different numerical resolution (to a lesser degree), the fraction of preprocessed satellites residing in low-mass hosts of $10^{13-13.5}\Ms$ in TNG300 is about 10 per cent higher with respect to TNG100.} galaxies falling as satellites (solid curves), as a function of $z=0$ galaxy stellar mass, stacked into four $z=0$ host mass bins, as labeled in the legend. The shaded areas denote the Poissonian error in each mass bin.
For completeness and to guide the eyes, we also show the fractions of galaxies directly falling as centrals into the main branch of their current host (dashed curves), these curves being complementary to the solid ones.

Overall, pre-processing in TNG is predominant in clusters ($\gtrsim 10^{14}\,\Ms$) compared to low-mass groups ($<10^{13.5}\,\Ms$), by about 20 percentage points at any stellar mass, and slightly decreases with increasing galaxy stellar mass. We ascribe this host-mass dependence of the fraction of pre-processed satellites to the nature of the hierarchical growth of structures.
About 65 (40) per cent of  $10^{9-9.5} \, \Ms$ satellites residing in hosts of mass $>10^{14.6} \, \Ms$ ($10^{13-13.5} \, \Ms$) have previously been part of groups of at least $10^{12} \, \Ms$ and then fallen into (the progenitor of) their final host.
For more massive satellites of $10^{11}\,\Ms$, this fraction drops to 50 per cent in massive clusters and to 20 per cent in low-mass groups.
These results are consistent with those by \cite{2013Bahe,2019Bahe}, which are also based on cosmological hydrodynamical simulations of large sets of hosts.

The bottom left panel of Fig. \ref{fig:qfrac_preproc} shows the average probability of TNG satellites having been pre-processed, i.e. orbiting in other previous hosts of at least $10^{12} ~ \Ms$, as a function of the time since last infall and as a function of $z=0$ galaxy stellar mass.
Interestingly, we find a non-monotonic correlation between the probability of being pre-processed and time since last infall, with the highest probability occurring for satellites that were accreted into their final host between 3 and 6 Gyr ago. As sensible, satellites that fell in more than 8 Gyr ago have a low probability of being pre-processed by other groups -- less than 30 per cent -- and this is also the case for satellites accreted between 1 and 2 Gyr ago.
Additionally, the few satellites that today are more massive than $10^{11.5}\, \Ms$ were almost certainly part of other subgroups, falling into their current host between 2 and 8 Gyr ago.

It is likely that the time the galaxies have spent in their preprocessing hosts and the mass of the preprocessing hosts are both key factors in determining how much the galaxy is affected by preprocessing \citep{2018Han}. Therefore, we show the distributions of both these properties in the bottom right panel of Fig. \ref{fig:qfrac_preproc}.
The color denotes the galaxy number density in the depicted plane while the contours encompass different fractions of satellites, divided into 2D bins of size 0.75 dex in maximum mass of the preprocessing host and 0.25 Gyr in time spent as preprocessed satellite. The running median is represented by a solid black line.

The maximum mass of the host is evaluated as the peak mass of the group in which a satellite resides, right before being captured by the $z=0$ one. 
Since pre-processed satellites may have been part of more than one previous host and since any such hosts will also increase in mass with time, we must choose a representative definition for the mass of the preprocessing hosts. We focus on the maximum mass of the host to gauge the maximal effect of pre-processing.

Finally, we show only those galaxies that remain satellites until $z=0$, never leaving their host, so that the y-axis is effectively the time spent in any host (of at least $10^{12} ~ \Ms$) before the $z=0$ one. Additionally, it is worth noticing that the y-axis of Fig. \ref{fig:qfrac_preproc} has to be intended as the ``total'' amount of time that a satellite may have spent in other hosts -- even if there were more than one such hosts.

As expected, Fig. \ref{fig:qfrac_preproc} shows that the more massive the host (peak mass) in which a satellite is preprocessed, the longer the time it spent there. Of more relevance is the fact that galaxies can be pre-processed for a long time: for as long as 3-4 Gyr on average in hosts more massive than $10^{14.5}\, \Ms$, and up to 5-6 Gyr in the tails of the distributions. On the other hand, satellites have spent on average less than 2 Gyr in pre-processing hosts of maximum mass $10^{12-13} ~\Ms$.

\subsection{Fractions of pre-processed and pre-infall quenched satellites}

In the top panel of Fig. \ref{fig:qfrac_preproc} we have demonstrated that pre-processing is more frequent among satellites currently residing in clusters than in groups.
The left panel of Fig.~\ref{fig:preproc_prequench} shows an integrated version of Fig. \ref{fig:qfrac_preproc}: the percentage of TNG\footnote{In all fractions presented here we bracket the outcome from TNG100 (leftmost numbers in each tile) and TNG300 (rightmost numbers in each tile). The differences between TNG100 and TNG300 are mostly driven by the fact that the two simulations represent the evolution of two different volumes, one larger than the other, and with different relative contributions to the galaxy populations of those residing in very massive hosts.} satellites residing in hosts of at least $10^{12} ~ \Ms$ before the current host, which is about 40 per cent of the whole population (central box) \citep[see also][for similar results]{2009Mcgee,2018Han}. We further divide the $z=0$ satellite sample into four tiles -- low and high-mass satellites residing in group or cluster hosts, all evaluated at $z=0$ -- and show the percentage of pre-processed galaxies in each of them. This picture clearly shows that pre-processing is more common in clusters than in groups, for both low and high-mass satellites, as shown above. Additionally, the frequency with which this occurs is higher for satellites below $10^{10} ~\Ms$, for which TNG returns about 54-57 per cent of pre-processed satellites residing today in clusters. 

The fractions of Fig.~\ref{fig:preproc_prequench}, left, do not tell us yet whether pre-processing has had an effect on the internal properties of satellite galaxies. Nevertheless, due to the high percentage of pre-processed satellites (Figs.~\ref{fig:qfrac_preproc} and \ref{fig:preproc_prequench}, left) and the long amounts of time that they might have spent in other subgroups (Fig.~\ref{fig:qfrac_preproc}, bottom), it is plausible that a non-negligible fraction of satellites fall into their current host already quenched and that such a fraction may be a significant contribution to the overall $z=0$ quenched fractions. 
We have discussed this phenomenon in Section \ref{intro} and already shown its impact on quenched fractions for TNG in Fig.~\ref{fig:frac_r2}. In Fig.~\ref{fig:preproc_prequench}, we explicitly provide the numbers of TNG satellites that were already quenched at the time of infall into in their current host -- either as centrals (before any infall) or as satellites -- normalized by the number of $z=0$ {\it quenched} satellites.

Among the $z=0$ quenched satellites, about 28 per cent are already quenched at the time of last infall. 
The percentages in the four tiles (representing the same satellites population of the left panel) demonstrate that $\sim$ 35-37 (30) per cent of satellites more massive than $10^{10} ~\Ms$ residing in clusters (groups) quench before falling into their current host, compared to $\sim$27-30 (11-16) per cent of low-mass satellites.
 
It is important to realize that the (higher) percentages of {\it massive} satellites that were already quenched before their last infall also include those that quenched when they were centrals, likely because of their AGN feedback, and not necessarily because of pre-processing, as already discussed in Section \ref{sec:qfrac}. Hence, in the next section we proceed to better investigate the diverse pathways towards the quenching of satellites, detailing what occurs at different stellar masses.

\subsection{The effect of pre-processing and internal processes on the $z=0$ quenched fractions of satellites}
\label{sec:contributions}

\begin{figure*}
\centering
\includegraphics[width=16cm]{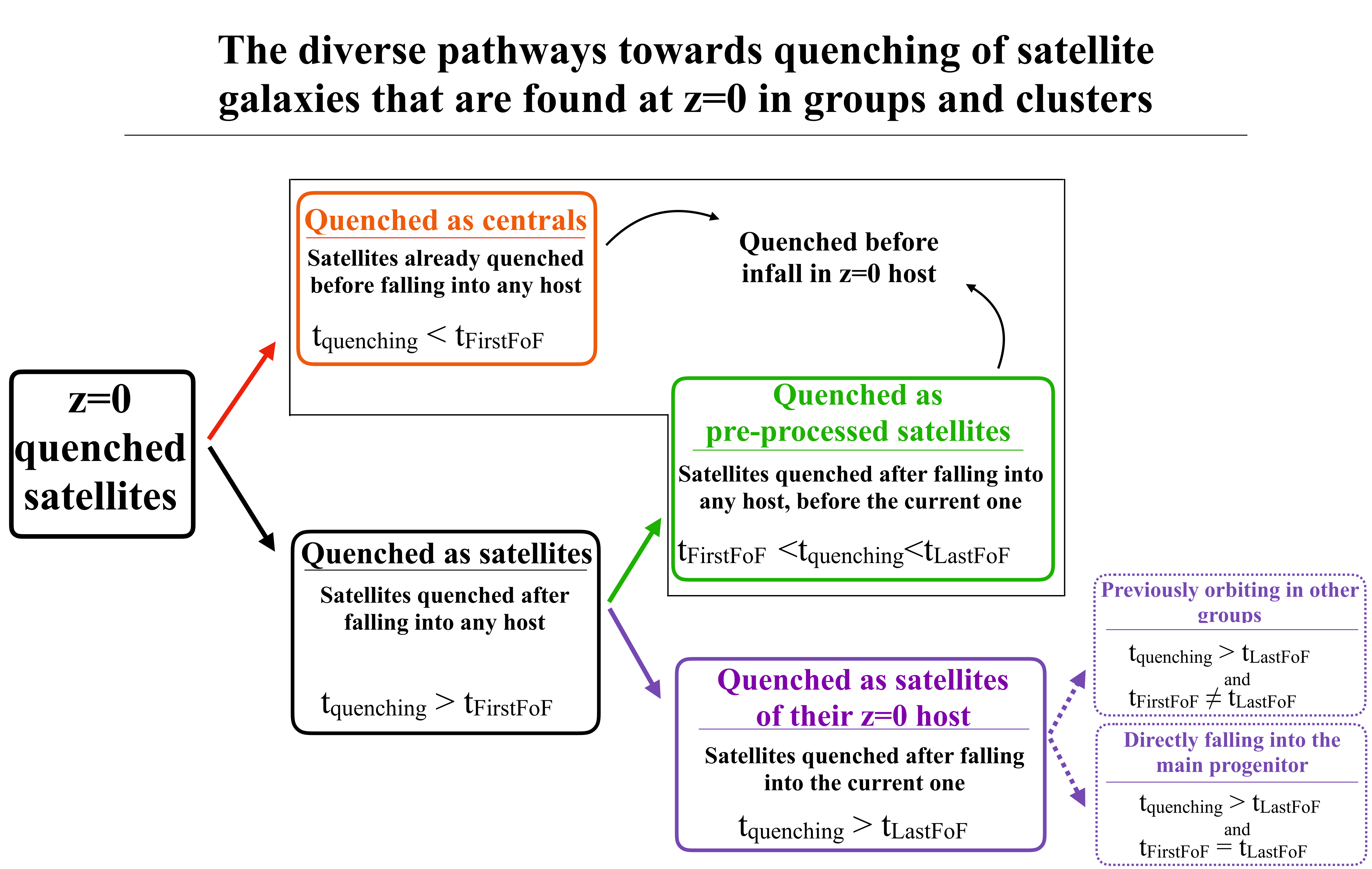}
\caption{\label{fig:quenched_population} 
{\bf The diverse pathways towards the quenching of satellite galaxies that are found at $z=0$ in groups and clusters}. Satellites residing today within groups and clusters may have followed different pathways to quenching. Some of them could have halted their star formation because of their AGN feedback, prior to their accretion into a dense environment (orange box). Others may have quenched after falling into a host, either within a pre-processing host (green box) or within the current one (purple box), the latter category also including satellites previously orbiting in other hosts (pre-processed but not quenched) or directly falling in the $z=0$ one. All these pathways are responsible for the trends in quenched fractions and magnitudes of e.g. Figs.~\ref{fig:Q_frac}, \ref{fig:environment}, \ref{fig:frac_r2}.
}
\end{figure*}

A detailed inspection of the histories of TNG galaxies reveals that satellites found at $z=0$ in groups and clusters may have followed diverse pathways to quenching: these pathways are schematically summarised in Fig.~\ref{fig:quenched_population}. 
 
Firstly, satellites can be quenched before any infall because of their internal processes ({\bf quenched as centrals}, orange box) or after being accreted into any hosts ({\bf quenched as satellites}, black box). Furthermore, the latter category can be further split into two, based on whether a satellite was quenched as it was orbiting in its current host ({\bf satellites quenched in their $z=0$ host}, purple box) or when it was a member of other pre-processing hosts ({\bf satellites quenched as pre-processed satellites}, green box). Satellites that either quenched as centrals or while being pre-processed are both part of the satellites that were quenched before their last infall into the $z=0$ host (as depicted by the common black box). Their frequency has been given and discussed with Fig.~\ref{fig:preproc_prequench}, right panel: about 30 per cent of the $z=0$ quenched population in the studied mass ranges. 
Additionally, even if a satellite was actually quenched in its $z=0$ host, this does not rule out the possibility that it was orbiting in a different host before infalling in the current host (purple dashed boxes) and that this pre-processing might have partially contributed to the quenching.

In Fig.~\ref{fig:quenched_population} we also detail the operational definitions adopted to practically assign $z=0$ satellites to these various categories. These are based on the comparison between the infall times, as explained in Section \ref{Infalltime} and Tab.~\ref{tab:infalltime}, and the quenching time. The latter is defined as the last time the satellite's SFR falls below 1 dex from the main sequence, evaluated using the procedure described in Section \ref{QvsSF}.

\begin{figure*}
\centering
\includegraphics[width=0.49\textwidth]{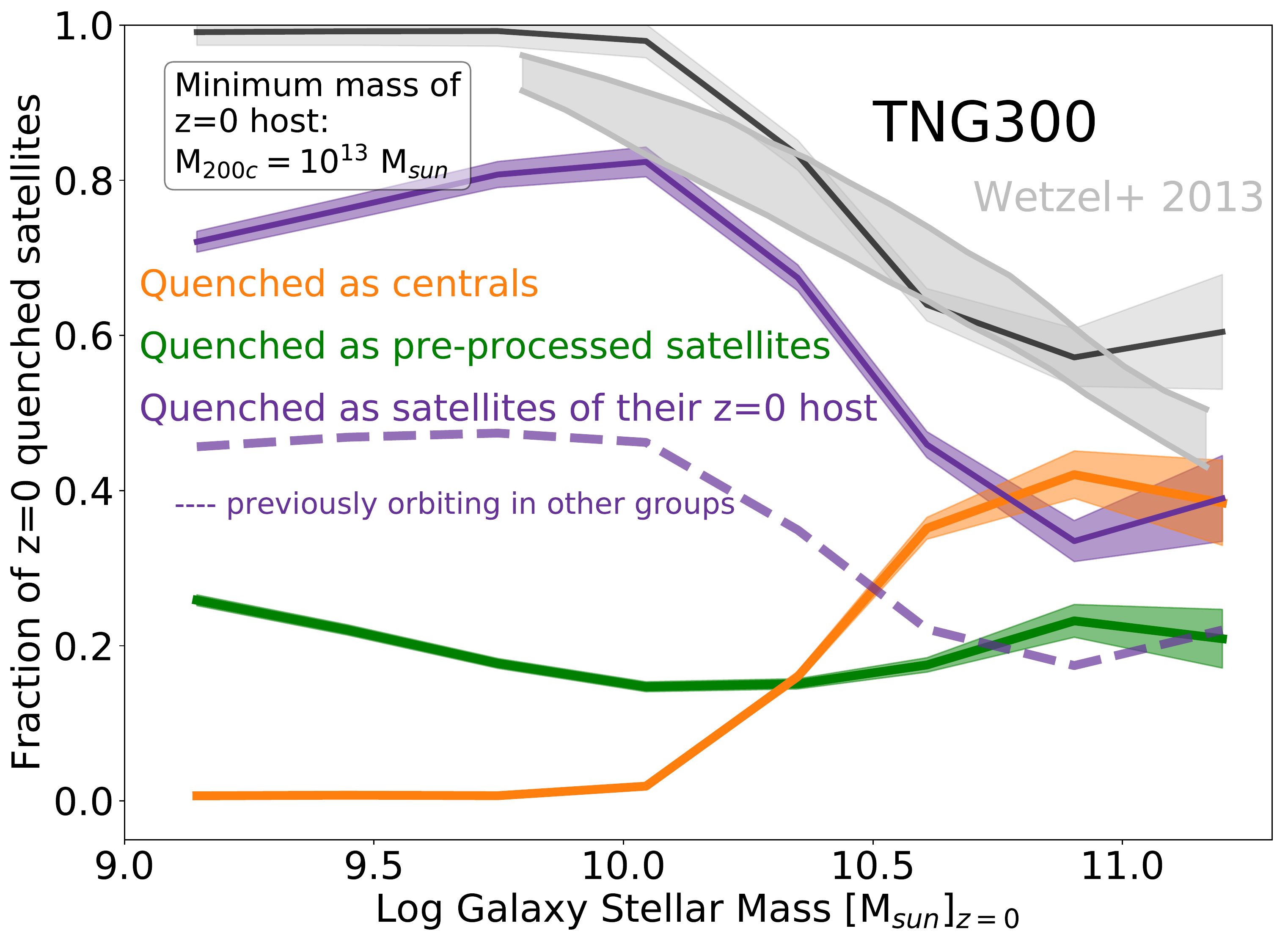}
\includegraphics[width=0.49\textwidth]{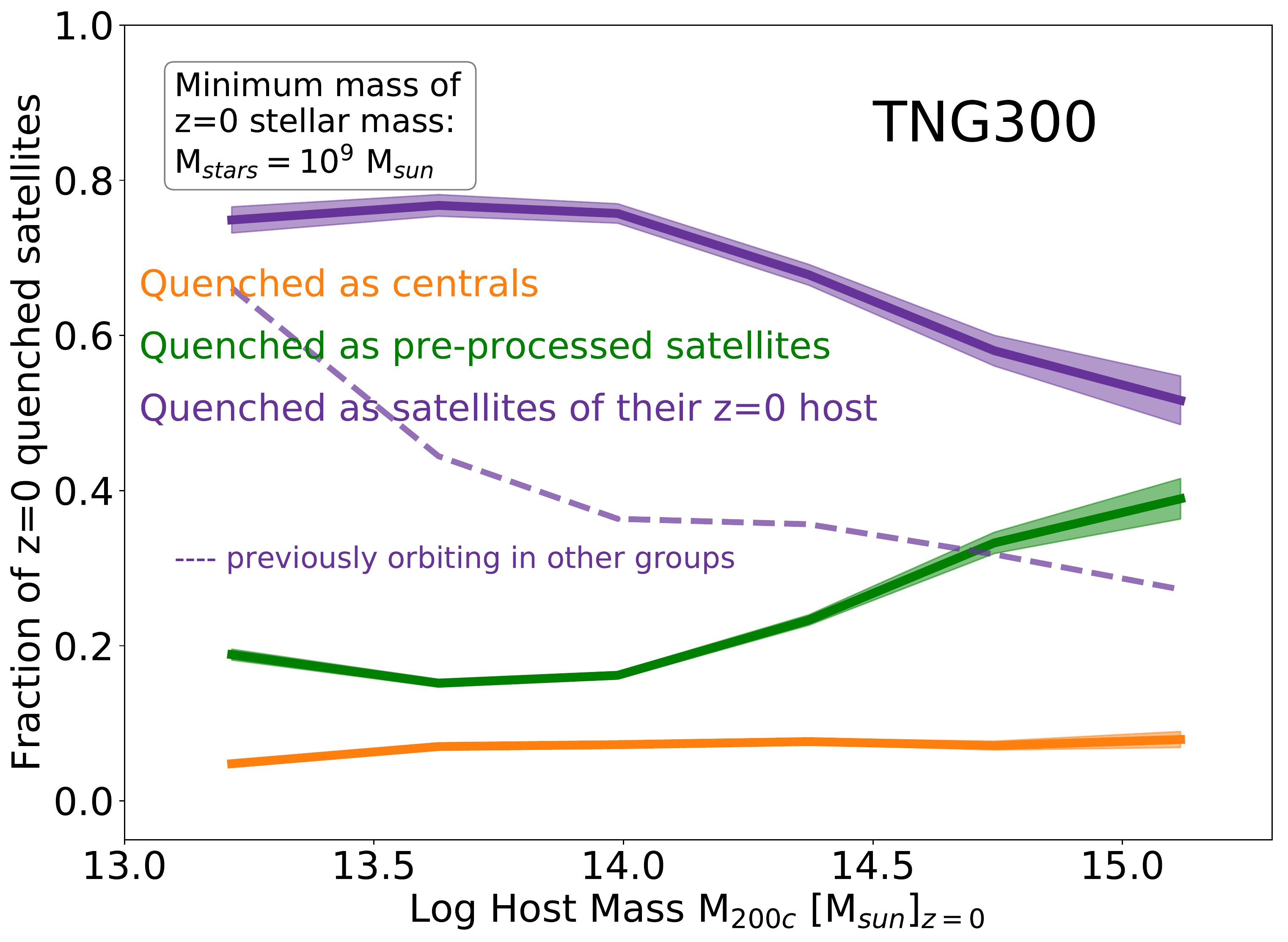}
\includegraphics[width=14cm]{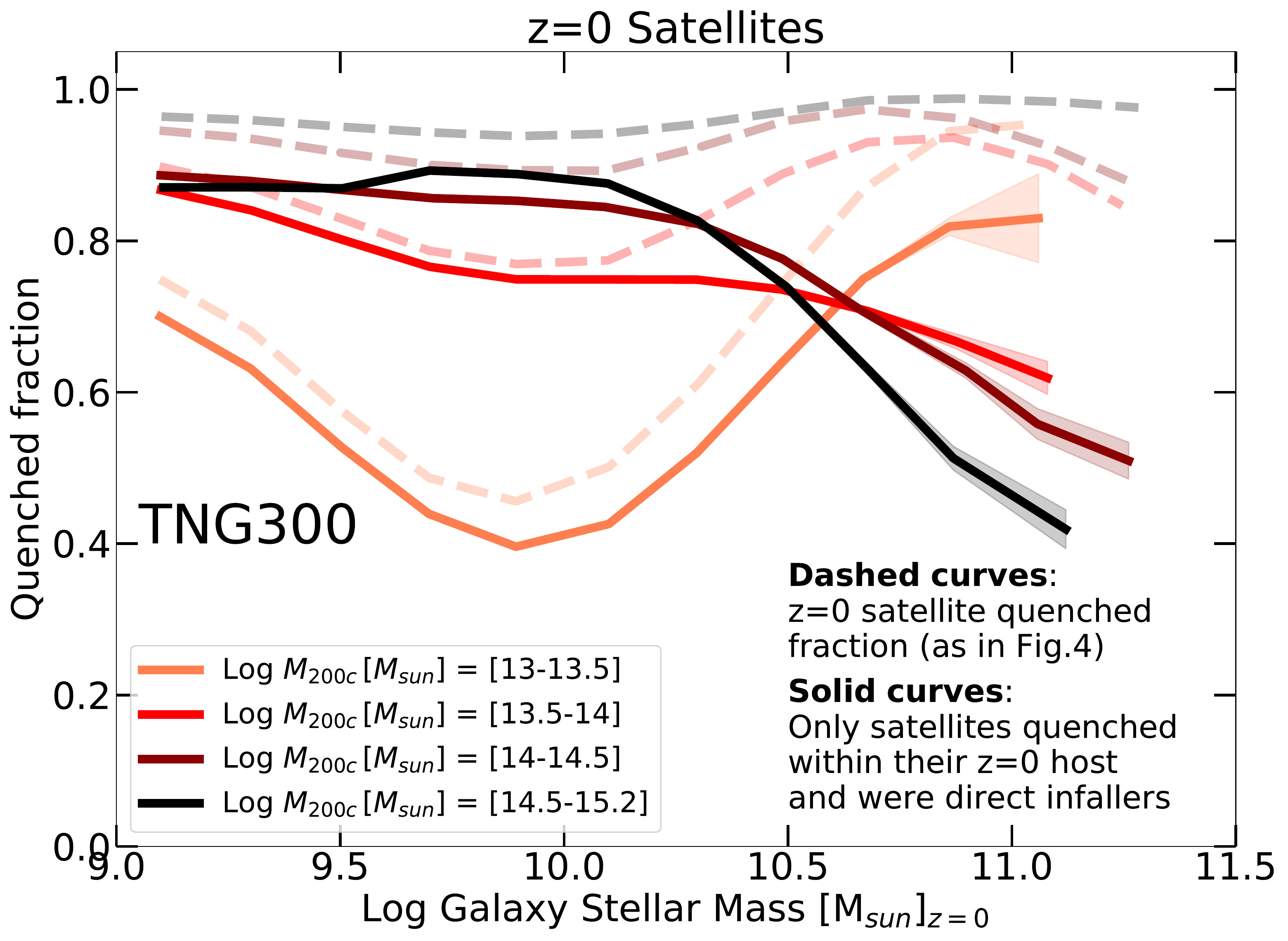}
\caption{\label{fig:pre-proc} 
Top: fraction of $z=0$ quenched satellites versus galaxy stellar mass (left panel) and versus host mass (right panel), both measured at $z=0$.
In both panels, colors denote the three different sub-samples that constitute the $z=0$ quenched population (see Fig. \ref{fig:quenched_population}): satellites quenched as centrals (orange curves), as pre-processed (green curves) and as satellites in their $z=0$ hosts (purple curves). The black curve represents satellites quenched as satellites, i.e. the sum of the green and purple curves. Additionally, we also show the fraction of galaxies quenched as satellites found by \citet{2013Wetzel} (grey shaded area), in broad agreement with TNG (see text for more details). The fraction is evaluated as the number of quenched satellites in each populations over the total number of quenched satellites in a specific mass bin. Bottom: quenched fraction of satellites that have fallen as centrals (direct infallers, solid curves) in the current host and were star-forming at accretion. For comparison, we also plot the left panel of Fig. \ref{fig:environment} as dashed curves.}
\end{figure*}

The top panels of Fig.~\ref{fig:pre-proc} show the fractions of $z=0$ quenched  satellites in TNG300 that were quenched via the different pathways summarized in Fig.~\ref{fig:quenched_population}, as a function of current galaxy stellar mass (left panel) and current host mass (right panel).
Even if not shown, the following results also hold for TNG100.
The fractions are evaluated as follows: in each mass bin of 0.2 dex, we count the number of galaxies quenched in one of the aforementioned scenarios and divide by the total number of $z=0$ quenched satellites in that bin. The shaded areas represent the Poissonian error and we only show mass bins with more than 10 satellites. 

We find that satellites more massive than a few $10^{10} ~\Ms$ today are the only ones that could quench when they were centrals (orange curve, left panel), i.e. before any infall: in TNG this is due to their AGN feedback and is in line with our findings already discussed in Section \ref{sec:qfrac}. At the very high-mass end ($\gtrsim 10^{10.5} ~\Ms$) this percentage is, on average, $\sim$40 per cent of the $z=0$ quenched population. Across the considered stellar mass range, this fraction is independent of the mass of the host where they are found today (orange curve, right panel).

On the other hand, quenched satellites that were quenched within their current host (purple curves) exhibit a strong correlation with galaxy stellar mass (left panel): the fraction is $\sim$ 80 per cent for low-mass satellites $\MS \lesssim 10^{10}~\Ms$ and decreases to $\sim$ 40 per cent for satellites more massive than $10^{10.5} ~\Ms$.

We omit showing a curve representing the fraction of all satellites that were quenched \textit{before} their last infall, for which we have provided and discussed the percentages in the right panel of Fig. \ref{fig:preproc_prequench}: this would be given by the sum of satellites quenched as centrals (orange curves) and of those quenched in any hosts other then their current ones (quenched as pre-processed, green curves).

The fraction of $z=0$ quenched satellites that quenched as satellites i.e. after any infall in any host is represented by the black curve: this is given by the sum of those quenched as pre-processed satellites (green curves) and those quenched as satellites of their z=0 host (purple curves). The gray shaded area in the left panel of Fig.~\ref{fig:pre-proc} represents the fraction of galaxies quenched as satellites found by \cite{2013Wetzel}, where the width indicates the uncertainty due to their adopted empirical parametrization for the evolution of SFRs of SDSS group/cluster satellites catalogues. 
Similarly to our analysis, the shaded area shows what fraction of currently quiescent satellites were quenched as satellites. This is in broad agreement with the black curve from TNG.

This analysis, overall, confirms two important results. 1) The build up of the $z=0$ quenched satellite population is strongly dominated at the low-mass end by the environmental effects imparted by group and cluster environments. 2) On the other hand, at the high stellar mass end ($\gtrsim 10^{10.5-11} ~\Ms$), quenching as a central (orange data) is almost as frequent as quenching as a satellite (purple + green data). It is important to note, however, that for the latter we have not yet conclusively demonstrated whether quenching occurs because of environmental processes alone, AGN feedback alone or, as is more plausible, a non-obvious interplay between the two -- we discuss this further in Section~\ref{sec:understanding}.

Finally, to specifically address the question of how important pre-processing is in quenching satellites, we emphasize that a) the majority of low-mass satellites that quench after (any) infall, seem to owe the halt of their star formation to their current host: purple versus green curves; and b) satellites quenched in any hosts other then their current ones (quenched as pre-processed, green curve, left panel) show an almost constant fraction of 20-30 per cent, regardless of galaxy stellar mass, for all hosts above $10^{13}$. This would seem to indicate that overall, pre-processing is unimportant, directly quenching at most only 1/3 of the low-mass $z=0$ quenched population of group and cluster satellites. There are two caveats to this general statement.

Firstly, even among those satellites quenched in their final host (purple curves), up to $\sim$50 have been pre-processed by subgroups (dashed purple curve, left panel), suggesting that, even if the hosts in which the satellites reside today are fundamental in modifying their star-formation activity, they may also have been helped by group pre-processing.

Furthermore, the fraction of satellites quenched in their current host decreases with increasing $z=0$ host mass (purple curve, right panel), thus ushering an intriguing possibility: even if the highest quenched fractions are found in massive clusters, this may not mean that they quench their satellites more efficiently compared to groups, once the role of pre-processing and quenching as centrals is taken into account.

Secondly, not only the frequency (Section~\ref{sec:preprocessing}), but also the relative importance of pre-processing for quenching is a strong function of current host mass: as shown in the right panel of Fig.~\ref{fig:pre-proc} where the statistics is dominated by low-mass galaxies, on average, about 50 per cent of the quenched satellites residing in clusters more massive than $10^{14.5} ~\Ms$ are actually quenched in these hosts (purple curve), but the remaining half (i.e. not just 20-30 per cent as for lower-mass hosts: green curve) become passive in other subgroups, entering the $z=0$ host already quenched. Additionally, among those satellites that were quenched in their $z=0$ host, roughly half were previously orbiting in other subgroups (dashed purple curve, right panel). These facts suggest that, if the quenched fractions of $10^{9.5-10} ~\Ms$ satellites of massive $10^{14.5-15} ~\Ms$ hosts at $z\sim0$ are higher in TNG than in the Universe (as we find by 10-20 percentage points when comparing to those inferred from SDSS: see  \textcolor{blue}{Donnari et al. 2020b)}, this may be due to a combination of phenomena, i.e. not only those pertaining to the environmental processes in very massive hosts at recent cosmic epochs. Future and detailed comparisons of satellite quenched fractions in group-mass hosts at intermediate and high redshifts will be of the essence to inform possible modeling issues.

\subsection{On the role of massive hosts and of AGN feedback in massive satellites}
\label{sec:understanding}

The knowledge from the previous section allows us to go back to the left panel of Fig. \ref{fig:environment}, in which we have shown the quenched fraction of $z=0$ satellites versus galaxy stellar mass, stacked in four bins of $z=0$ host mass, to highlight how this fraction is shaped within (and because of) different environments. We are now able to answer the question: how does this picture change when pre-processing is accounted for or, conversely, once we exclude satellites quenched before the last infall into the $z=0$ host?

In order to better isolate the efficiency of hosts as a function of host mass at quenching their satellites, we show in the bottom panel of Fig.~\ref{fig:pre-proc} the quenched fraction evaluated as the number of satellites falling as centrals (and then not-preprocessed) that were quenched in their $z=0$ hosts over the total number of satellites that fell as centrals and that were still star-forming at the time of accretion (solid curves), thereby excluding satellites quenched because of pre-processing or quenched prior to infall because of internal processes.
For reference, we also plot the curves from the left panel of Fig. \ref{fig:environment}, in dashed. 
Two key results can be seen in this figure. For satellites with masses below $10^{10.5} ~ \Ms$, more massive clusters are {indeed} more efficient at quenching than lower mass hosts, particularly with respect to lower-mass groups of $10^{13-13.5} ~ \Ms$: black and brown curves versus orange curve.
However, we have to bear in mind that this result is convolved with the distributions of satellite infall times, which depend on host mass \citep[see e.g.][their Fig.7]{2020Engler}, and we have seen in Fig.~\ref{fig:frac_r2} that longer infall times imply higher probabilities of being quenched. We have checked (even if not shown) that the trend with host mass of Fig.~\ref{fig:pre-proc}, bottom, would appear stronger (i.e. more massive clusters would appear progressively more efficient at quenching than lower-mass hosts) once we restrict to sub-populations of satellites with similar infall times, particularly with comparable recent infall times of 1 to 4 billion years ago.

On the other hand, at the high-mass end of the satellite population that was still star-forming at accretion, an inversion of the trend with host mass appears, whereby the quenched fractions are {\it lower} in clusters than in groups, by up to 40 percentage points. This is extremely intriguing: this inversion is inconsistent with quenching being driven by environment, as environmental processes are expected to manifest themselves as more being effective at quenching in more massive hosts -- see previous results.

The ``host-rank'' inversion of Fig.~\ref{fig:pre-proc}, together with the absence of a cluster-centric distance dependence of the quenched fraction of massive satellites (Fig.~\ref{fig:frac_r2}, middle right panel), imply that the main culprits for quenching massive satellites are in fact secular and internal processes, these being AGN feedback in TNG. Furthermore, such processes are made {\it less} effective in higher mass hosts. In \citealt{2020Joshi}, we have shown that, in TNG, gas accretion into black holes (BH) is significantly suppressed for satellite galaxies after their accretion into $\gtrsim10^{14}~ \Ms$ in comparison to galaxies of similar mass ``in the field'' (their Fig. 14). Lower BH accretion in turn would reduce the overall energy injected as BH feedback throughout the lifetime of massive satellites, ultimately reducing or delaying its quenching effects\footnote{Because of the numerical implementation of BH positionining in the TNG model \citep{2017Weinberger}, a fraction of satellite galaxies may not have a super massive BH. Even if we do not show it here for the sake of brevity, we have checked and we think that the `host-rank' inversion of Fig.~\ref{fig:pre-proc} is not related to the lack of BHs is some satellites. For satellites more massive than $\sim 10^{10.5} \Ms$, the fraction of those with no BHs is small (less than 10 per cent) and has a little-to-no dependence on host mass in between $10^{13-14.5} \Ms$. At higher host masses, this fraction grows to about 20 per cent, but it remains subdominant and thus we think it cannot be responsible for the host dependencies we find in the bottom panel of Fig. \ref{fig:pre-proc}.}.

The phenomenology described here and predicted by the TNG simulations, whereby AGN feedback is hindered for satellites in massive environments, is hard to prove with observations because of the degeneracies of the quenching pathways described so far; but it would be very interesting to see whether it is common to other galaxy formation models. In fact, it is important to bear in mind the following: asserting that quenching in massive satellites is driven by internal processes does not exclude the possibility of witnessing signatures of environmental processes, such as ram-pressure stripped gaseous tails, which in fact are clearly visible in massive TNG satellites across cosmic epochs \citep{2019Yun}. Moreover, a long-lasting and population-averaged environmental obstruction to AGN activity is not necessarily in contrast to previous findings which have shown that higher density cluster environments can actually trigger enhanced BH activity in satellites for a period of time, which in some instances can in turn act to eventually quench the satellites \citep[e.g.][]{2013Mcgee, 2017Poggianti,2020Ricarte}.

\begin{figure*}
\centering
\includegraphics[width=16cm]{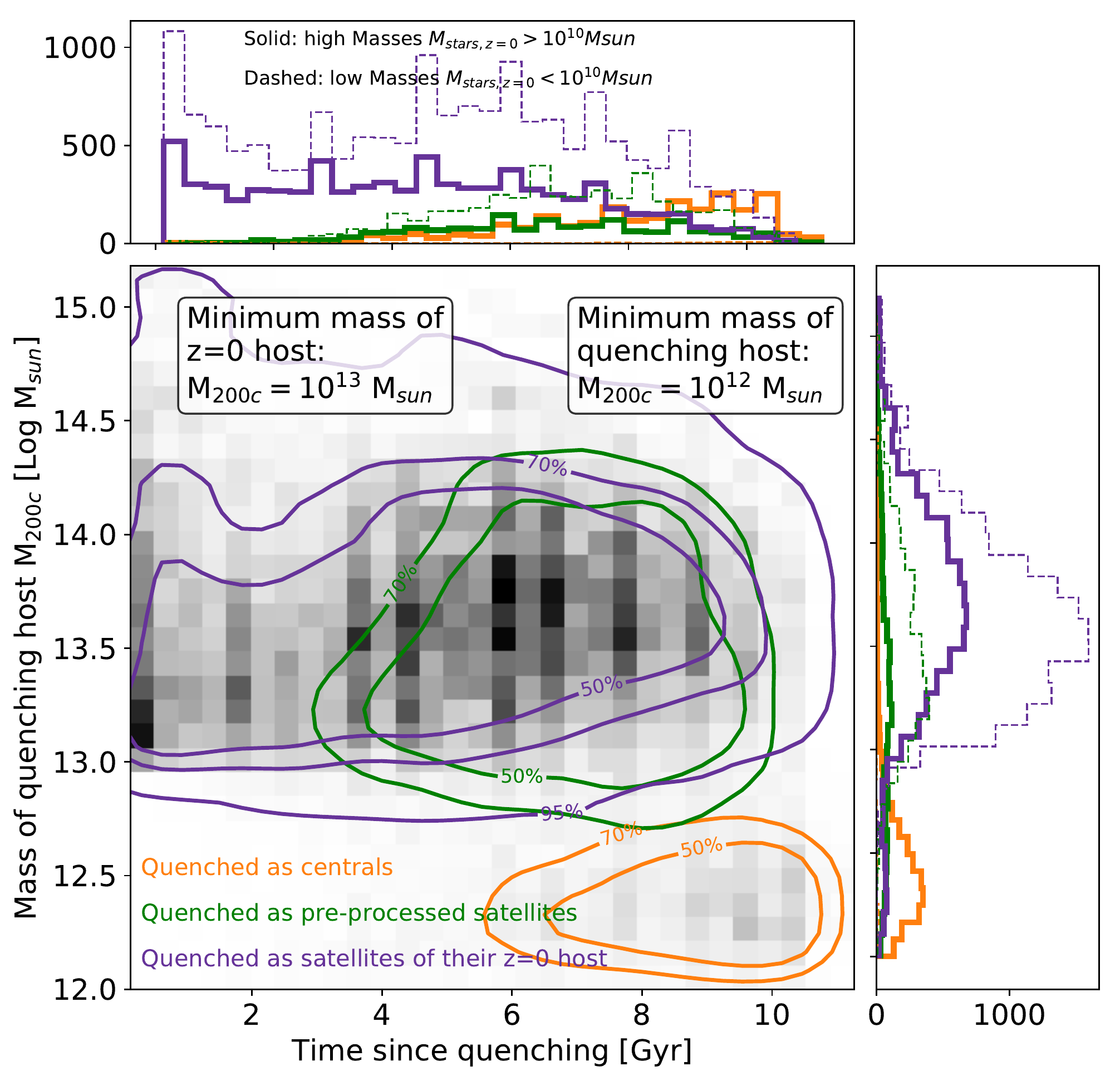}
\caption{\label{fig:Histo_Mhost_quenching} 
{\bf Distribution of the mass of quenching hosts and the time since quenching of $z=0$ group and cluster satellites}.
Mass of quenching host versus time of quenching (lookback time) for the three subsamples discussed in the text: 
satellites already quenched as centrals (orange), those quenched as pre-processed satellites (green) and those quenched in their current host (purple). Contours represent 50, 70 and 95 per cent of each subsample, while shades of grey denote the number of galaxies in each pixel. 
The majority of satellites that quenched in their $z=0$ host became passive in groups of $\simeq10^{13.5-13.8}~ \Ms$ between 0 and 10 Gyr ago. Only satellites more massive than $10^{10} \, \Ms$ quenched as centrals $\sim$ 9 Gyr ago, likely due to secular internal processes.}
\end{figure*}
\subsection{Distribution of the quenching times and the host masses of quenching hosts}
We conclude our analysis by exploring the typical mass scale of hosts where satellites quench and how long ago they actually quenched.

In Fig. \ref{fig:Histo_Mhost_quenching}, main panel, we plot the 2D histogram of the mass of the ``quenching'' host (i.e. the mass of the halo host at the time of quenching) against the time since quenching, dividing our sample in 40x40 bins on the plane.
The color is the number density of $z=0$ satellites and the colored contours encompass 50 and 70 per cent of satellites quenched as centrals (orange), those quenched as pre-processed satellites (green) and those quenched in their current host (purple), for which we include also the contour encompassing 95 per cent. In the main panel, all $z=0$ satellites are considered, with stellar mass $\geq10^9~\Ms$ in hosts with total mass exceeding $10^{13}~\Ms$.

The top and right histograms represent the distributions of the quenching time and the host mass at quenching, respectively, with the same color code as of the central panel. Additionally, we show results for high-mass ($>10^{10}~\Ms$, solid curves) and low-mass satellites ($<10^{10}~\Ms$, dashed curves) separately.

Satellites quenched prior to their last infall, but after being accreted by a dense environment (pre-processed satellites, green curves), typically became passive between 6 to 8 Gyr ago, the majority of them in groups of $M_{200c}=10^{13-14} ~\Ms$ (at the time of quenching), with no strong dependence on $z=0$ stellar mass. 

Satellites quenched within their $z=0$ host (purple curves), instead, exhibit a broader distribution of quenching times -- between 1 and 7 Gyr ago -- regardless of their $z=0$ stellar mass. Because of this, the typical mass scale of hosts in which they were actually quenched can be lower, since the progenitors of their $z=0$ hosts would have been less massive than the ones in which they currently reside. In fact, we find a narrow distribution around $M_{200c}=10^{13.2-14} ~\Ms$ (between 16th and 84th percentiles).
So overall, across the whole $z=0$ satellite population that quenched as satellites, quenching most frequently occurs in hosts of mass $10^{13-14}\, \Ms$ over the last many billion years of cosmic evolution.

Finally, as already shown in Fig. \ref{fig:pre-proc}, only satellites more massive than $10^{10-10.5} \, \Ms$ quench as centrals (orange curves), before falling into any host. In their case, the label``quenching host" may be misleading, as in fact the depicted mass scale refers to their own underlying dark matter haloes. Satellites that quenched as centrals typically quenched in a narrow halo mass range, at $10^{12.1-12.6} ~\Ms$ (between 16th and 84th percentiles). For massive satellite galaxies that are now found in groups and clusters and quenched as centrals, quenching typically occurred between 6 and 10 Gyr ago (between 16th and 84th percentiles).

\section{Summary and conclusions}
\label{summary}

In this paper, we have investigated the star formation activity of $\MS = 10^{9-12}~\Ms$ galaxies using two cosmological
hydrodynamical simulations from the IllustrisTNG project, TNG100 and TNG300. In particular, we have focused on group and cluster satellites, i.e. galaxies found today within the virial radius of haloes with total mass $M_{200c}=10^{13-15.2}~\Ms$. 
We have explored how the quenched fractions differ for central versus satellite galaxies and how they change as a function of galaxy stellar mass, redshift, host mass, cluster-centric distance, and time since infall. 

We find that the IllustrisTNG model returns the qualitative features of the two main quenching scenarios widely recognized in both observations and previous theoretical models, namely environmentally-driven and mass quenching.
While we refer the reader to the companion paper, \cite{Donnari2020b}, for a detailed quantitative comparison to observational results, here we have leveraged the outcome of the IllustrisTNG simulations to draw a complete picture of the diverse pathways that galaxies can take towards quenching within the hierarchical growth of structure scenario, and to put forward a plausible framework for the interpretation of observational findings. In particular, we have singled out the role that pre-processing and internal secular mechanisms (chiefly, AGN feedback) play in shaping the $z=0$ quenched fraction of satellite populations.\\

Our key findings are the following:
\begin{enumerate}
    \item Low-mass galaxies ($\MS \lesssim 10^{10}\Ms$) are rarely quenched (5-8 per cent for centrals) unless they are satellites of groups or clusters, consistently with the environmentally-driven quenching scenario. In the case of satellites, at fixed galaxy stellar mass, the fraction of low-mass passive satellites is higher in more massive hosts, closer to the host center, and for those accreted earlier into their current host, with an upper limit of about 4-6 billion years for the time needed for processes in groups and clusters more massive than $10^{13.5} \Ms$ to have an effect and hence for quenching to occur.\\
    
    \item Conversely, high-mass galaxies ($\gtrsim$ a few $ 10^{10}\Ms$) quench on their own: internal processes, i.e. AGN feedback in the case of IllustrisTNG, and not environment, is the main culprit for the cessation of star formation in massive galaxies, {\it regardless of whether they are centrals or satellites}. The massive satellite quenched fractions are high (80-90 per cent) regardless of host mass, cosmic time ($z\lesssim0.5$), cluster-centric distance, and time since infall.\\
    
    \item A non-negligible subset (about 30 per cent) of the quenched satellite populations of groups and clusters at low-redshift were already quenched before infalling into their last host: for low-mass satellites, this is entirely due to environmental effects during pre-processing, i.e. while orbiting in smaller hosts prior to infall into their final one; for high-mass satellites, quenching occurred when they were centrals, because of their AGN feedback.\\
    
    \item Both environmental and internal star-formation quenching can occur at early cosmic epochs, when the Universe was only a couple billion years old. While, as expected, the quenched fractions of IllustrisTNG central and satellite galaxies are generally lower at higher redshifts (Fig.~\ref{fig:frac_redshift}), frequent manifestations of environmental processes in hosts more massive than about $10^{13}\,\Ms$ are already in place at $z\lesssim 1$ and the bulk of the $z=0$ group and cluster quenched satellites ceased their star formation many billion years ago: 6-10, 1-7, and 4-8 Gyrs ago (16th-84th percentiles) for satellites that quenched as centrals, in their current host, or as satellites before falling into it, respectively. 

\end{enumerate}

Overall, the IllustrisTNG simulations reveal a great diversity of quenching pathways for satellite galaxies that are found today within groups and clusters of galaxies (see schematic in Fig.~\ref{fig:quenched_population}). We have demonstrated that satellites can quench before any infall in dense environments  (i.e. in our jargon, as centrals) or after being accreted into any hosts (i.e. quenched as satellites) and that among the latter, satellites can quench while being members of pre-processing hosts other than the one where they are found today (see schematic in Fig.~\ref{fig:sketch}). 
The frequency of each quenching pathway and their contribution to the $z=0$ satellite quenched fraction depend strongly on the mass of both satellites and hosts under consideration at the time of inquiry (or observation). Therefore, comparisons among models and observational inferences cannot be performed without taking this into account and different satellite and host mass ranges or different satellite and host mass distributions may be partially the cause for the apparently contradicting results in the literature. 

Importantly, not only the frequency, but also the relative impact of pre-processing on quenching is a function of current host mass. About 40-60 per cent of satellites that are found today in massive clusters ($M_{200c}\gtrsim10^{14-14.5}~\Ms$) have been members of other subgroups of at least $M_{200c}=10^{12}~\Ms$ before falling into the one in which we find them today (i.e. have been pre-processed). These compare to 30-50 per cent of pre-processed satellites that are currently members of lower-mass groups ($M_{200c}\sim 10^{13-14}~\Ms$ -- Fig. \ref{fig:qfrac_preproc} and left panel of Fig. \ref{fig:preproc_prequench}). Furthermore, members of today's massive clusters have been pre-processed for as long as 3-4 billion years on average and up to 5-6 billion years in the tails of the distributions (Fig.~\ref{fig:qfrac_preproc}, bottom right panel). It is therefore not surprising that pre-processing is responsible for the quenching of about half of the low-mass satellites that inhabit the most massive clusters today ($M_{200c}\gtrsim10^{14.5-15.2}~\Ms$) and not just one third as when averaged across the group and cluster host mass function (Fig.~\ref{fig:quenched_population}, top right panel).

The typical and most frequent host mass where environmental quenching takes place is $M_{200c} = 10^{13-14}~\Ms$, by averaging across the $z=0$ satellite and host populations and accounting for the last many billion years of cosmic evolution: this finding once more confirms the role of groups (rather than clusters) as fundamental building blocks of the hierarchical cosmic structure formation. Still, after excluding satellites quenched by their AGN feedback and those that have been quenched by pre-processing, massive clusters remain more efficient at quenching their members compared to lower-mass hosts: this is the case for satellites with stellar masses below about $10^{10.5} ~\Ms$ (Fig. \ref{fig:pre-proc}) and is in line with the widely-accepted dependence of environmental quenching processes such as ram-pressure stripping on environmental density and depth of the host gravitational potential well. 

On the other hand, for the more massive satellites, the IllustrisTNG model predicts an inversion of the ``ranking of host efficiency'', in the sense that the percentage of massive satellites that actually quench in their current host is higher in groups rather than in clusters. According to IllustrisTNG, AGN feedback is hindered for satellites in dense environments or, in other terms, massive hosts reduce on average the effectiveness of quenching internal secular processes. While observationally arduous to be proven, it will be interesting to see whether a long-lasting and population-averaged environmental obstruction to AGN activity is common also to other galaxy formation models.

\section*{ Data availability}
The data from the Illustris and TNG simulations used in this work are publicly available at the websites \url{https://www.illustris-project.org} and \url{https://www.tng-project.org}, respectively \citep{2015Nelson_release, 2019Nelson_release}.

\section*{Aknowledgement}
The authors are grateful to the referee for their comments and suggestions that helped to improve an earlier version of this manuscript.
MD acknowledges support from the Deutsche Forschungsgemeinschaft (DFG, German Research Foundation) -- Project-ID 138713538 -- SFB 881 (``The Milky Way System'', subproject A01).
FM acknowledges support through the Program "Rita Levi Montalcini" of the Italian MIUR.
MV acknowledges support through an MIT RSC award, a Kavli Research Investment Fund, NASA ATP grant NNX17AG29G, and NSF grants AST-1814053, AST-1814259 and AST-1909831.
The flagship simulations of the IllustrisTNG
project used in this work have been run on the HazelHen Cray
XC40-system at the High Performance Computing Center Stuttgart
as part of project GCS-ILLU of the Gauss centres for Super-computing(GCS). Ancillary and test runs of the project were also run on the Stampede supercomputer at TACC/XSEDE (allocation
AST140063), at the Hydra and Draco supercomputers at the Max
Planck Computing and Data Facility, and on the MIT/Harvard computing facilities supported by FAS and MIT MKI.

\footnotesize{
\bibliographystyle{mn2e}
\bibliography{biblio}}

\appendix

\section{Effects of numerical resolution on quenched fractions of centrals and satellites}
\label{appendix}

\begin{figure*}
\centering
\includegraphics[width=0.49\textwidth]{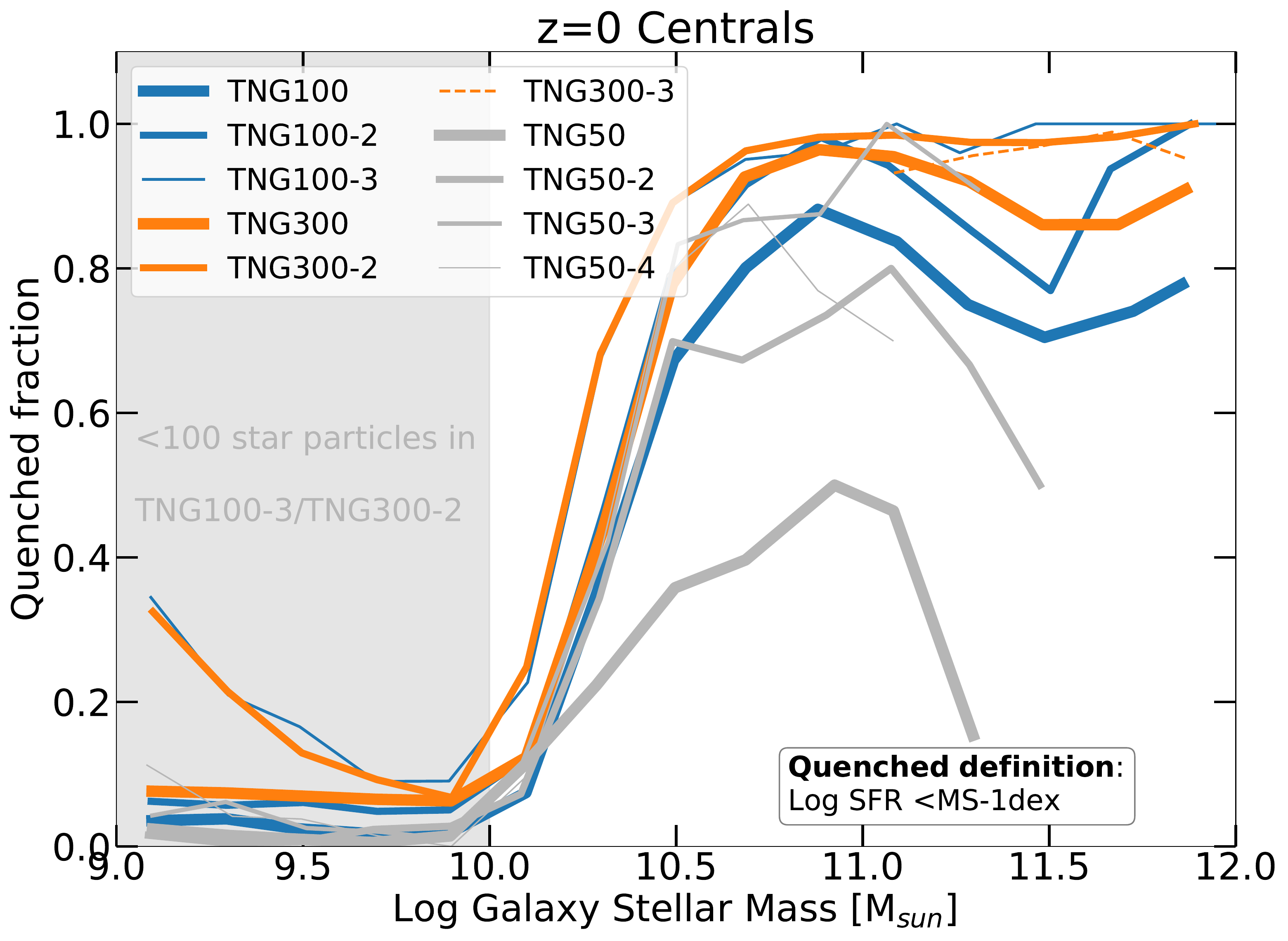}
\includegraphics[width=0.49\textwidth]{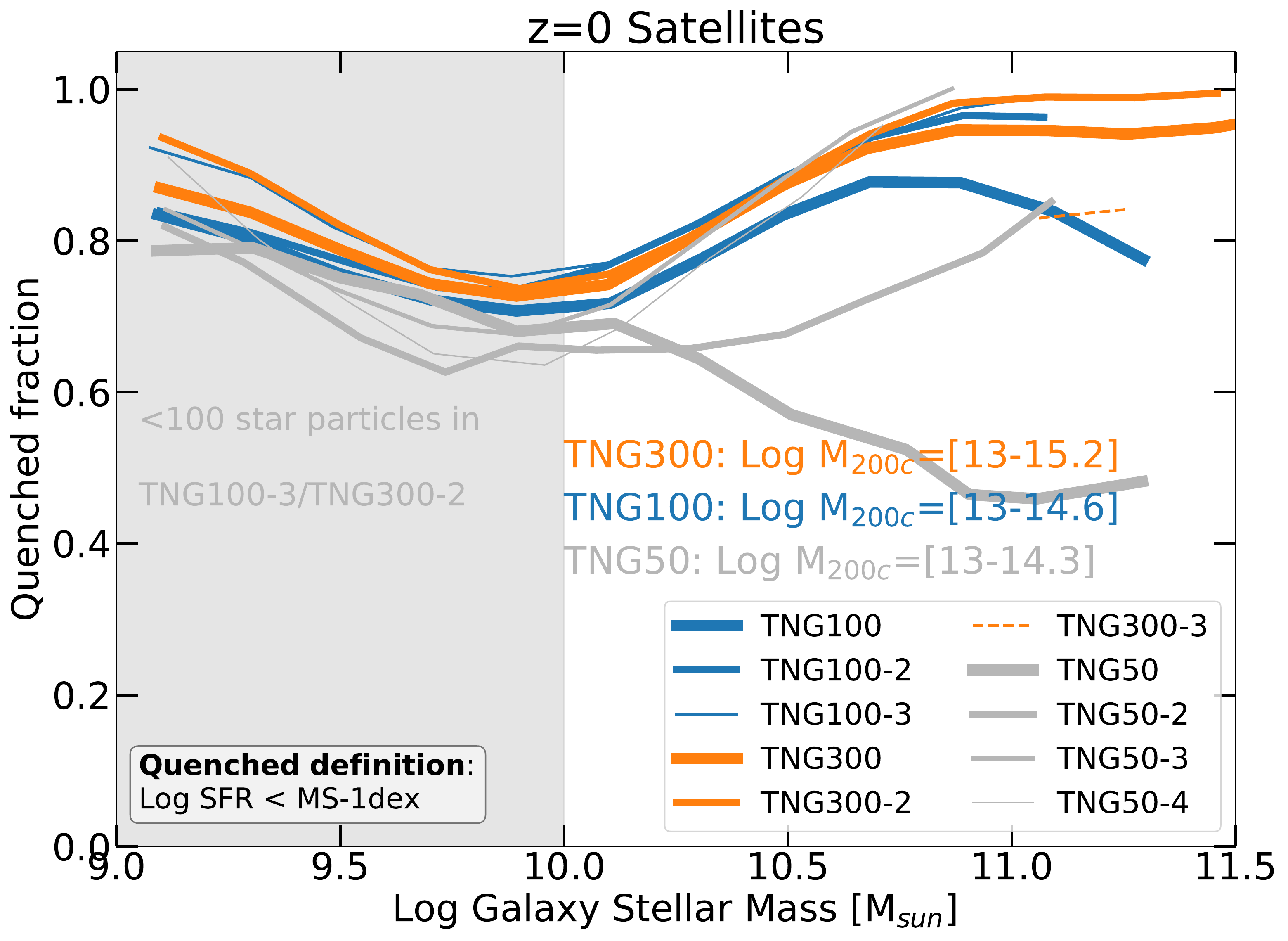}
\includegraphics[width=0.49\textwidth]{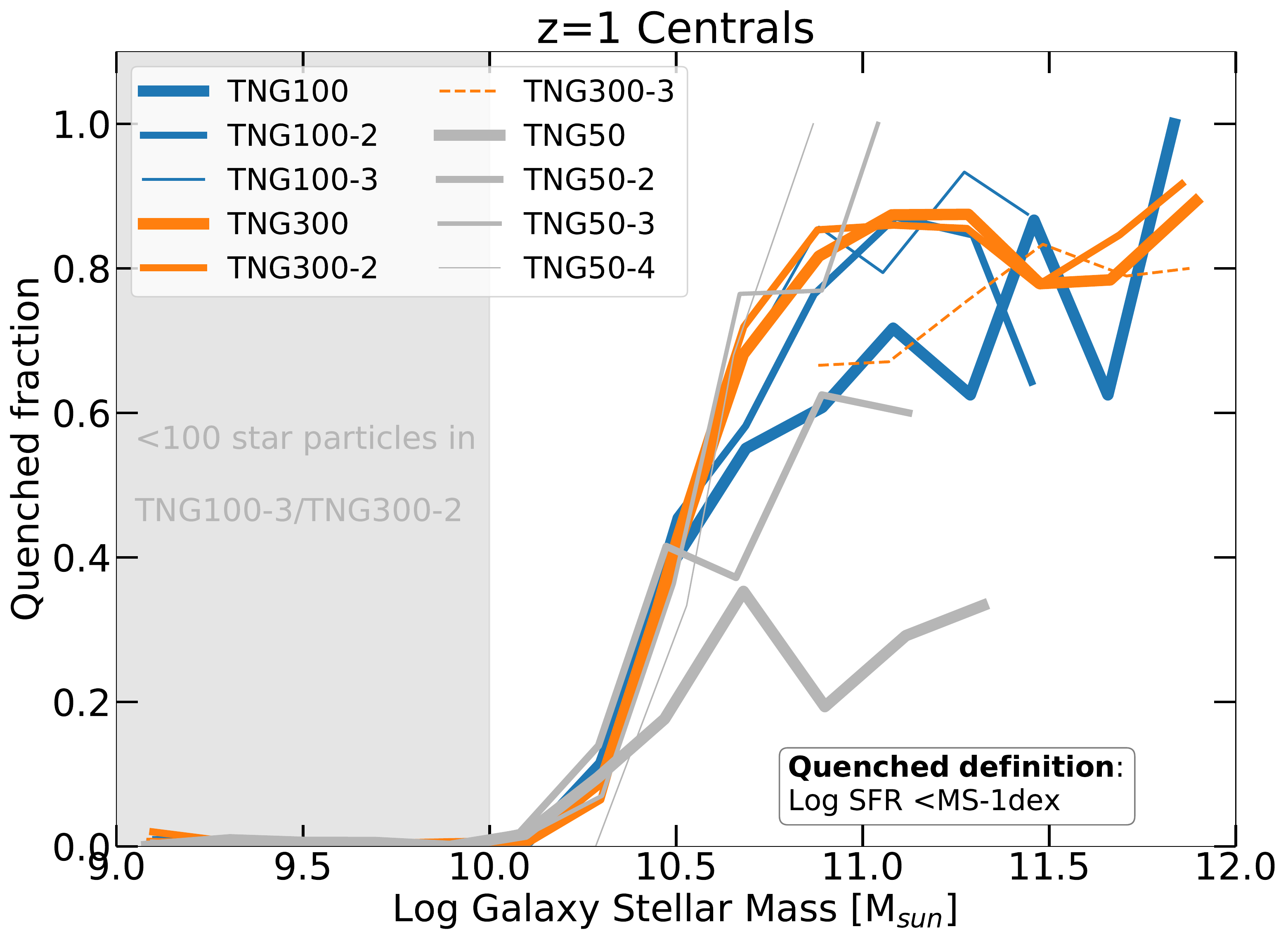}
\includegraphics[width=0.49\textwidth]{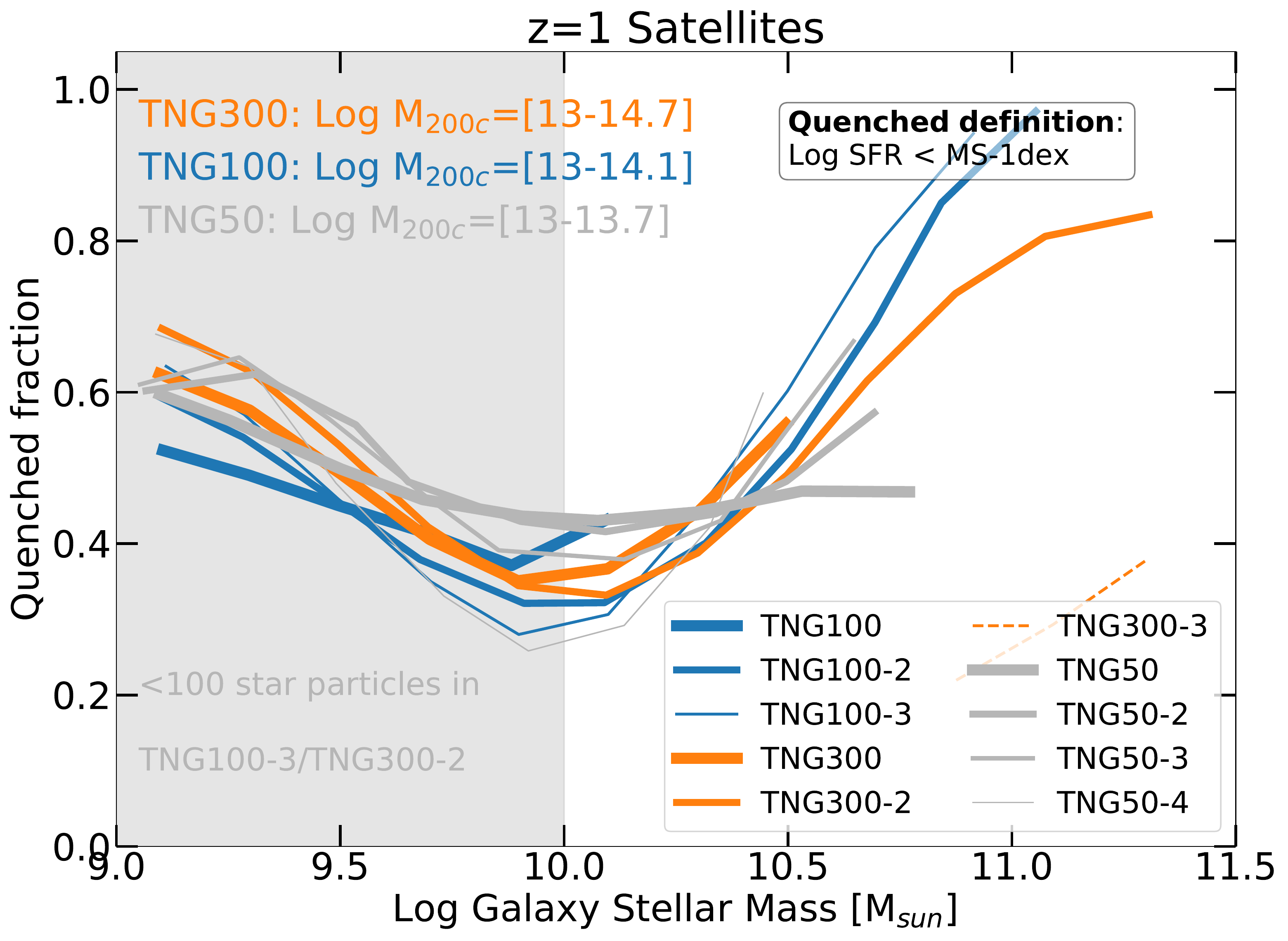}
\caption{\label{fig:resolution} {\bf Resolution study of the quenched fractions}. Fraction of quenched galaxies as a function of stellar mass for centrals (left) and satellites (right) at $z=0$ (top) and $z=1$ (bottom) according to the three different resolution levels of TNG300 (orange) and TNG100 (blue), and the four resolution levels of TNG50 (grey). The grey shaded area denotes the resolution limit of TNG100-3 and TNG300-2, where the sample of centrals includes galaxies with less than one hundred of stellar particles.}
\end{figure*}

In this Appendix, we quantify the effect of numerical resolution on quenched fractions. Necessarily, the comparisons in this Appendix will also be affected by sample variance, as -- as noted throughout the text -- TNG100 and TNG300 span different simulation box sizes.

Here, we show results according to TNG100 and TNG300 at their different resolution levels, namely TNG100-2, TNG100-3 and TNG300-2, TNG300-3, respectively. In the TNG100 series, the baryonic mass reads $1.4\times 10^6, 1.1\times 10^{7},$ and $8.9\times 10^{7} \, \Ms$, respectively, while in the TNG300 series we have $1.1\times 10^7, 8.8\times 10^{7}$, and $7.0\times 10^{8} \, \Ms$ \citep[see also Tab. 1 in][ for more details]{2018Pillepich}, so that the resolutions of e.g. TNG300 and TNG100-2 are identical.
For completeness, and albeit not used for this work, we also include in this Appendix the results from the highest-resolution run of the IllustrisTNG series, TNG50, with its lower-resolution realizations: TNG50-2, TNG50-3 and TNG50-4. The different resolution realizations of TNG50 are characterized by a baryonic mass of $8.5\times 10^4 \Ms$, $6.8\times 10^5 \Ms$, $5.4\times 10^6 \Ms$, and $4.3\times 10^7 \Ms$, respectively \citep[see e.g][]{2019Pillepich_50}, so that that TNG100 has a mass resolution in between TNG50-2 and TNG50-3, while TNG100-2 (and TNG300) have a mass resolution in between TNG50-3 and TNG50-4. TNG50 encompasses a smaller volume than TNG100 and TNG300 (about 50 comoving Mpc a side, in comparison to about 100 and 300 comoving Mpc of TNG100 and TNG300): it therefore lacks the highest-mass end clusters (above a few 10$^{14}\, \Ms$) that this work focuses on.

Fig. \ref{fig:resolution} shows the analog of Fig. \ref{fig:Q_frac}. We compare quenched fractions of centrals (left panel) and satellites (right panel) at $z=0$ (top) and $z=1$ (bottom) according to the three resolution levels of TNG300 and TNG100, represented with different curve thickness. Satellites in the right columns include those in hosts with minimum masses of $10^{13} \Ms$ and maximum masses labeled in each panel, depending on the run and redshift.
To avoid confusion, results from TNG300-3 are presented only above its mass resolution where the samples of centrals and satellites include galaxies with at least one hundred stellar particles, namely $\MS\gtrsim10^{11} \, \Ms$ (dashed curve).
Moreover, shaded gray area indicate the stellar mass ranges where galaxies are resolved with fewer than 1000 stellar particles in the TNG100-3 and TNG300-2 runs, which have the same numerical resolution.

Typically, the lower resolution runs return larger quenched fractions, both for centrals and satellites, at any stellar masses, with no noticeable redshift trend (at least at $z\le1$), but with some inversions of trends for more massive satellites at high redshifts. Focusing on TNG100 and TNG300 and barring the outcome of the lowest resolution run of the series that returns completely disparate results (TNG300-3, with a baryonic mass resolution of $7.0\times 10^{8} \, \Ms$), the resolution-induced discrepancy across factors of 8 in particle mass resolution is however rather small, being typically smaller than about 10 percentage points at $z\le1$, but being somewhat confounded by sampling and statistics, and therefore reaching larger deviations up to 15-20 percentage points in regimes with poor sampling, like at the highest-mass end at higher redshifts. This indicates that at fixed galaxy stellar mass the TNG300/TNG100 results carry a systematic average error of about 10 percentage points across the regimes studied in this paper: this is comparable or even smaller than the differences between inherent simulation results versus observation-mocks needed for comparison to observational results \citep[see][]{Donnari2020b}. However, within the IllustrisTNG framework and as further demonstrated and discussed by \citealt{Donnari2020b}, the higher-resolution run TNG50 (not used in this paper) returns $z=0$ quenched fractions that are up to another 10-30 percentage points lower than those from TNG100: as we discuss there, and given that TNG50 simulates fewer high-mass galaxies and hosts due its smaller comoving volume, this is due to a combination of resolution effects and sample variance. 
Nonetheless, Fig.~\ref{fig:resolution} confirms what was already pointed out in Section \ref{sec:qfrac}: the differences between TNG100 and TNG300, albeit small, are mainly due to a numerical resolution effect. However, higher numerical resolutions and poorer sampling statistics can impact the predictions of the quenched fractions with systematic uncertainties by up to 10-40 percentage points, depending on galaxy mass, host mass, and redshift \citep[see also][]{Donnari2020b}.

\label{lastpage}
\end{document}